\documentclass[nojss]{jss}

\usepackage{thumbpdf,lmodern}
\usepackage{verbatim}
\usepackage{bm}
\usepackage{amsfonts}
\usepackage{pgf, tikz}
\usepackage{booktabs}
\usepackage{xcolor}
\usepackage{comment}
\usetikzlibrary{arrows,automata,fit}
\usepackage{subfig}
\usepgflibrary{shapes.geometric}
\newtheorem{definition}{Definition}
\usepackage[bottom]{footmisc}
\usepackage{framed}
\usepackage{multirow}
\usepackage{natbib}
\usepackage[binary-units = true]{siunitx}

\author{Federico Carli\\ Universit\`{a} degli Studi di Genova
\And  Manuele Leonelli\\ IE University
\AND Eva Riccomagno \\ Universit\`{a} degli Studi di Genova
\And Gherardo Varando\\ Universitat de Val\`{e}ncia}
\Plainauthor{Federico Carli, Manuele Leonelli, Eva Riccomagno, Gherardo Varando}

\title{The \proglang{R} Package \pkg{stagedtrees} for Structural Learning of Stratified Staged Trees}
\Plaintitle{The R Package stagedtrees for Structural Learning of Stratified Staged Trees}
\Shorttitle{The \proglang{R} Package \pkg{stagedtrees}}
\Abstract{
  \pkg{stagedtrees} is an \proglang{R} package which includes several algorithms
for learning the structure of staged trees and chain event graphs from data. Score-based and clustering-based algorithms are implemented, as well as various functionalities to plot the models and perform inference. The capabilities of \pkg{stagedtrees} are illustrated using mainly two datasets both included in the package or bundled in \proglang{R}.
}

\Keywords{chain event graphs, graphical models, \proglang{R}, staged trees, structure learning algorithms}
\Plainkeywords{chain event graphs, graphical models, R, staged trees, structure learning algorithms}

\Address{
Gherardo Varando\\
Image Processing Laboratory (IPL) \\
Parc Cient\'{i}fic Universitat de Val\`{e}ncia\\
C/ Catedr\'{a}tico Jos\'{e} Beltr\'{a}n, 2 \\
46980 Paterna (Val\`{e}ncia). Spain \\
E-mail: \email{gherardo.varando@uv.es}
}

\begin{document}



\section[Introduction]{Introduction} \label{sec:intro}

In the past twenty years there has been an explosion of the use of graphical models to represent the relationship between a vector of random variables and perform distributed inference which takes advantage of the underlying graphical representations. Bayesian networks (BNs) \citep{Darwiche2009,Fenton2012} are nowadays the most used graphical models, with applications to a wide array of domains and implementation in various software: for instance, the \proglang{R} packages \citep{R} \pkg{bnlearn} by \citet{Scutari2010} and \pkg{gRain} by \citet{Hojsgaard2012}, among others.

However, BNs can only represent symmetric conditional independences which in practical applications may not be fully justified. For this reason, a variety of models that can take into account the asymmetric nature of real-world data have been proposed; for example, context-specific BNs \citep{Boutilier1996}, labeled directed acyclic graphs \citep{Pensar2015} and probabilistic decision graphs \citep{Jaeger2006}. Unlike most of its competitors, the chain event graph (CEG) \citep{Collazo2018,Smith2008,riccomagno2004identifying,riccomagno2009geometry} can capture all (context-specific) conditional independences in a unique graph, obtained by a coalescence over the vertices of an appropriately constructed probability tree, called staged tree.

CEGs have been used for cohort studies \citep{Barclay2013}, causal analysis \citep{Thwaites2010,thwaites2013causal} and case-control studies \citep{keeble2017adaptation,keeble2017learning}. Structure learning algorithms have been defined in the literature \citep{barclay2014chain,Collazo2016,Silander2013,cowell2014causal}. The user's toolbox to efficiently and effectively perform uncertainty reasoning with CEGs further includes methods for inference and probability propagation \citep{Gorgen2015,Thwaites2008}, the exploration of equivalence classes \citep{Gorgen2018} and robustness studies \citep{Leonelli2019,Wilkerson2019}. The model class of CEGs and staged trees have been further extended to model dynamic problems with recursively updated probabilities \citep{Barclay2015,Freeman2011a}, decision problems under the framework expected utility maximization \citep{Thwaites2017} and Bayesian games \citep{Thwaites2018}.

The \proglang{R} package \pkg{stagedtrees} implements some algorithms for learning staged trees and CEGs from data and is freely available from the Comprehensive R Archive Network (CRAN) at \url{http://CRAN.R-project.org/package=stagedtrees}. The package also provides inferential and visualization functions for such models as well as descriptive and summary statistics about the graph structure. The only other software available to learn such models is the \pkg{ceg} package \citep{Collazo2017}, including one learning algorithm \citep[\textit{Agglomerative Hierarchical Clustering},][]{Freeman2011}. 

\section[Staged trees and chain event graphs]{Staged trees and chain event graphs} \label{sec:staged}

Many statistical graphical models represent a random vector of interest in terms of undirected or directed acyclic graphs. In particular, BNs are directed acyclic graphs where each vertex corresponds to a random variable and a missing edge between two nodes represents conditional independence. Conversely, staged trees are directed trees equipped with probabilites where atomic events coincide with root-to-leaf paths.

A directed tree $\mathcal{T}=(V,E)$ is a tree with vertex set $V$ and edge set $E$, where each vertex except for the root has one parent only, all non-leaf vertices have at least two children and all edges point away from the root. For $ v,v' \in V $ let $e = (v, v') \in E$ be the edge pointing from $v$ to $v'$. For a non-leaf $v$, let $E(v) = \{v'\in V: (v,v')\in E\}$ and call $\mathcal{F}(v) = (v, E(v))$ a floret of the tree. Let $\Theta$ be a non-empty set of labels and $\theta: E \rightarrow \Theta$ be a function such that for any non-leaf $v \in V$ the labels in $\theta(E(v))$ are all distinct. The set $\theta(E(v))$ is denoted by $\bm{\theta}_v$ and is called the set of floret labels. Next assume $\Theta\subseteq [0,1]$. If $\sum_{e\in E(v)}\theta(e)=1$ for all non-leaf $v$, then $\mathcal{T}$ together with the $\bm{\theta}_v$'s is called a \textit{probability tree} and $\theta(e)$ is the probability of the edge $e \in E$. Each root-to-leaf path $\lambda$ in $\mathcal{T}$, equivalently each leaf vertex, is associated to an atom in a discrete probability space and the atomic probabilities can be defined as $\prod_{e\in\lambda}\theta(e)$. Throughout, edges on a root-to-leaf path $\lambda$ are ordered from the closest to the root to the closest to the leaf. The atomic probabilities together with $\Theta$ give the statistical model associated to the tree.
\begin{definition}
A probability tree where for some $v,v'\in V$ $\bm\theta_v=\bm\theta_v'$, is called a \emph{staged tree}. The vertices $v$ and $v'$ are said to be in the same stage.
\end{definition}
Although not strictly required, a probability tree can represent the joint probability distribution of a discrete random vector $\bm{X}=(X_1,\dots,X_n)$ taking values in a product space $\mathbb{X}=\times_{i=1}^n\mathbb{X}_i$, where $\mathbb{X}_i$ is the finite sample space of $X_i$, $i = 1, \ldots, n$.

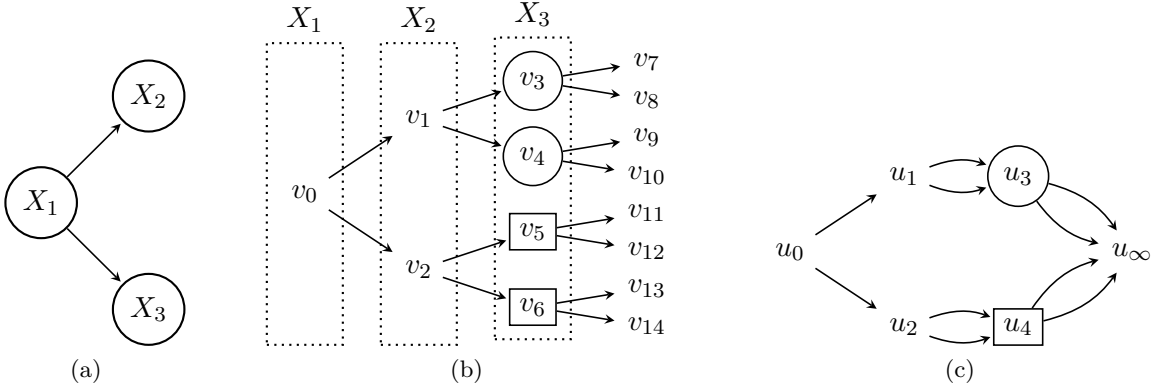
\begin{figure}
\centering
\subfloat[ \label{fig:bnex}]{
\begin{tikzpicture}[
            > = stealth, 
            shorten > = 1pt, 
            auto,
            node distance = 2cm, 
            semithick 
        ]

        \tikzstyle{every state}=[
            draw = black,
            thick,
            fill = white,
            minimum size = 8mm
        ]

        \node[state] (X1) {$X_1$};
        \node[state] (X2) [above right of=X1] {$X_2$};
        \node[state] (X3) [below right of=X1] {$X_3$};

        \path[->] (X1) edge   (X2);
         \path[->] (X1) edge   (X3);
 \end{tikzpicture}}
  \hspace{0.8cm}
\subfloat[ \label{fig:stagedtree}]{ 
\begin{tikzpicture}[
            > = stealth, 
            shorten > = 1pt, 
            auto,
            node distance = 3cm, 
            semithick 
        ]

        \tikzstyle{every state}=[
            draw = black,
            thick,
            fill = white,
            minimum size = 8mm
        ]
   \tikzstyle{frame} = [thick, draw=black, dotted,inner sep=0em,minimum height=4cm,minimum width=1cm]     
        
 \node (v0) at (0,0) {$v_0$}; 
 \node (v1) at (1.5,1) {$v_1$};
 \node (v2) at (1.5,-1) {$v_2$}; 
 \node[circle,draw] (v3) at (3,1.5) {$v_3$};
 \node[circle,draw] (v4) at (3,0.5) {$v_4$}; 
 \node[rectangle,draw] (v5) at (3,-0.5) {$v_5$};
 \node[rectangle,draw] (v6) at (3,-1.5) {$v_6$}; 
 \node (v7) at (4.5,1.75) {$v_7$};
 \node (v8) at (4.5,1.25) {$v_8$};
 \node (v9) at (4.5,0.75) {$v_9$};
 \node (v10) at (4.5,0.25) {$v_{10}$}; 
 \node (v11) at (4.5,-0.25) {$v_{11}$};
 \node (v12) at (4.5,-0.75) {$v_{12}$}; 
 \node (v13) at (4.5,-1.25) {$v_{13}$};
 \node (v14) at (4.5,-1.75) {$v_{14}$};
  \node[frame,fit=(v0),label=$X_1$]{};
  \node[frame,fit=(v1) (v2),label=$X_2$]{};
    \node[frame,fit=(v3) (v6),label=$X_3$]{};
  
  \path[->] (v0) edge   (v1);
  \path[->] (v0) edge   (v2);
  \path[->] (v1) edge   (v3);
  \path[->] (v1) edge   (v4);
  \path[->] (v2) edge   (v5);
  \path[->] (v2) edge   (v6);
  \path[->] (v3) edge   (v7);
  \path[->] (v3) edge   (v8);
  \path[->] (v4) edge   (v9);
  \path[->] (v4) edge   (v10);
  \path[->] (v5) edge   (v11);
  \path[->] (v5) edge   (v12);
  \path[->] (v6) edge   (v13);
  \path[->] (v6) edge   (v14); 
 \end{tikzpicture}
}
\hspace{0.8cm}
\subfloat[ \label{fig:CEG}]{ 
\begin{tikzpicture}[
            > = stealth, 
            shorten > = 1pt, 
            auto,
            node distance = 3cm, 
            semithick 
        ]
            \tikzstyle{every state}=[
            draw = black,
            thick,
            fill = white,
            minimum size = 8mm
        ]

 \node (u0) at (0,0) {$u_0$}; 
 \node (u1) at (1.5,1) {$u_1$};
 \node (u2) at (1.5,-1) {$u_2$}; 
 \node[circle,draw] (u3) at (3,1) {$u_3$};
 \node[rectangle,draw] (u4) at (3,-1) {$u_4$};
 \node (u5) at (4.5,0) {$u_{\infty}$};
  
  \path[->] (u0) edge   (u1);
  \path[->] (u0) edge   (u2);
  \path[->] (u1) edge [bend left=20]  (u3);
  \path[->] (u1) edge [bend right=20] (u3);
  \path[->] (u2) edge  [bend left=20] (u4);
  \path[->] (u2) edge  [bend right=20] (u4);
    \path[->] (u3) edge  [bend left=20] (u5);
  \path[->] (u3) edge   [bend right=20]  (u5);
  \path[->] (u4) edge  [bend left=20]  (u5);
  \path[->] (u4) edge   [bend right=20]  (u5);
 \end{tikzpicture}
}

\caption{Illustration of the construction of staged tree and CEG from a BN for three binary random variables. The BN in Figure \ref{fig:bnex} is represented by the $\bm{X}$-compatible tree in Figure \ref{fig:stagedtree} where the edges emanating from $v_0$ represent the outcomes of $X_1$; the edges emanating from $v_1$ and $v_2$ represent the outcomes of $X_2$ conditionally on the outcome of $X_1$; the edges emanating from $v_3,\dots,v_6$ represent the outcomes of $X_3$ conditionally on $X_1$ and $X_2$. The conditional independence of the BN coincides with the staging $\{v_3,v_4\}$ and $\{v_5,v_6\}$ (vertices not framed are in their own stage). The staged tree in Figure \ref{fig:stagedtree} is transformed into the CEG in Figure \ref{fig:CEG} using the positions $u_0=\{v_0\}$, $u_1=\{v_1\}$, $u_2=\{v_2\}$, $u_3=\{v_3,v_4\}$, $u_4=\{v_5,v_6\}$ and $u_\infty=\{v_7,\dots,v_{14}\}$. \label{fig:example}}
\end{figure}

Recall that for $\bm{x} = (x_1, \dots, x_n) \in \mathbb{X}$ the joint probability can be factorized according to the chain rule of probabilities 
\begin{equation}
\label{eq:chain}
p(\bm{x})=\prod_{i=2}^n ~ p(x_i|\bm{x}^{i-1}) ~ p(x_1),
\end{equation}
where $\bm{x}^{i-1}=(x_1,\dots,x_{i-1})\in\times_{j=1}^{i-1}\mathbb{X}_j$. This sequential factorization can be represented by a probability tree as the one in Figure \ref{fig:stagedtree} where the probabilities on the right-hand-side of Equation~(\ref{eq:chain}) are associated to the edges emanating from the non-leaf vertices. 

\vspace{0.2cm}

\begin{definition}
A probability tree $\mathcal{T}$ is called $\bm{X}$-compatible if for each $ \hspace{0.1cm} \bm{x} \in \mathbb{X}$ there exists a unique root-to-leaf path $ \lambda = (e_1, \dots, e_n)$  such that
$
\theta(e_{1}) = p(x_1)  
$ and
$
\theta(e_{i}) = p(x_i | \bm{x}^{i-1}) \hspace{0.2cm} for \hspace{0.1cm}  i = 2, \dots, n 
$. 
\end{definition}

An $\bm{X}$-compatible tree has as many leaves as elements in $\mathbb{X}$. All vertices at distance $i$ from the root are associated to the same random variable $X_{i+1}$, $i = 1, \dots, n-1$, and are said to be in the same \textit{stratum}.

Conditional independence statements embedded in BNs then correspond to equalities between probabilities on the right-hand-side of Equation~(\ref{eq:chain}). This can be captured in probability trees by identifying some of the floret probability values. 

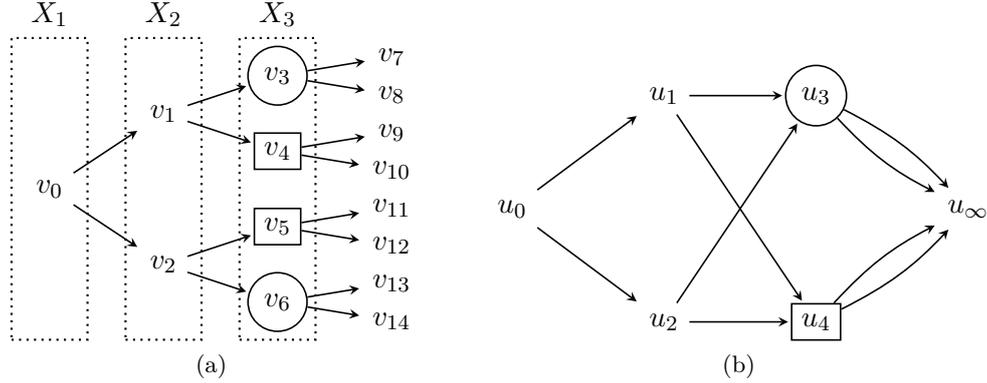
\begin{figure}
\centering
\subfloat[ \label{fig:stagedtree_2}]{ 
\begin{tikzpicture}[
            > = stealth, 
            shorten > = 1pt, 
            auto,
            node distance = 3cm, 
            semithick 
        ]

        \tikzstyle{every state}=[
            draw = black,
            thick,
            fill = white,
            minimum size = 8mm
        ]
   \tikzstyle{frame} = [thick, draw=black, dotted,inner sep=0em,minimum height=4cm,minimum width=1cm]     
        
 \node (v0) at (0,0) {$v_0$}; 
 \node (v1) at (1.5,1) {$v_1$};
 \node (v2) at (1.5,-1) {$v_2$}; 
 \node[circle,draw] (v3) at (3,1.5) {$v_3$};
 \node[rectangle,draw] (v4) at (3,0.5) {$v_4$}; 
 \node[rectangle,draw] (v5) at (3,-0.5) {$v_5$};
 \node[circle,draw] (v6) at (3,-1.5) {$v_6$}; 
 \node (v7) at (4.5,1.75) {$v_7$};
 \node (v8) at (4.5,1.25) {$v_8$};
 \node (v9) at (4.5,0.75) {$v_9$};
 \node (v10) at (4.5,0.25) {$v_{10}$}; 
 \node (v11) at (4.5,-0.25) {$v_{11}$};
 \node (v12) at (4.5,-0.75) {$v_{12}$}; 
 \node (v13) at (4.5,-1.25) {$v_{13}$};
 \node (v14) at (4.5,-1.75) {$v_{14}$};
  \node[frame,fit=(v0),label=$X_1$]{};
  \node[frame,fit=(v1) (v2),label=$X_2$]{};
    \node[frame,fit=(v3) (v6),label=$X_3$]{};
  
  \path[->] (v0) edge   (v1);
  \path[->] (v0) edge   (v2);
  \path[->] (v1) edge   (v3);
  \path[->] (v1) edge   (v4);
  \path[->] (v2) edge   (v5);
  \path[->] (v2) edge   (v6);
  \path[->] (v3) edge   (v7);
  \path[->] (v3) edge   (v8);
  \path[->] (v4) edge   (v9);
  \path[->] (v4) edge   (v10);
  \path[->] (v5) edge   (v11);
  \path[->] (v5) edge   (v12);
  \path[->] (v6) edge   (v13);
  \path[->] (v6) edge   (v14); 
 \end{tikzpicture}
}
\hspace{0.5cm}
\subfloat[ \label{fig:CEG_2}]{ 
\begin{tikzpicture}[
            > = stealth, 
            shorten > = 1pt, 
            auto,
            node distance = 3cm, 
            semithick 
        ]
            \tikzstyle{every state}=[
            draw = black,
            thick,
            fill = white,
            minimum size = 8mm
        ]

 \node (u0) at (0,0) {$u_0$}; 
 \node (u1) at (2,1.5) {$u_1$};
 \node (u2) at (2,-1.5) {$u_2$}; 
 \node[circle,draw] (u3) at (4,1.5) {$u_3$};
 \node[rectangle,draw] (u4) at (4,-1.5) {$u_4$};
 \node (u5) at (6,0) {$u_{\infty}$};
  
  \path[->] (u0) edge   (u1);
  \path[->] (u0) edge   (u2);
  \path[->] (u1) edge   (u3);
  \path[->] (u1) edge   (u4);
  \path[->] (u2) edge   (u4);
  \path[->] (u2) edge   (u3);
    \path[->] (u3) edge  [bend left=10] (u5);
  \path[->] (u3) edge   [bend right=10]  (u5);
  \path[->] (u4) edge  [bend left=10]  (u5);
  \path[->] (u4) edge   [bend right=10]  (u5);
 \end{tikzpicture}
}

\caption{Staged tree and CEG for three binary random variables with stages $\{v_0\}, \{v_1\}, \{v_2\}, \{v_3,v_6\}, \{v_4,v_5\}$ and positions $u_0=\{v_0\}$, $u_1=\{v_1\}$, $u_2=\{v_2\}$, $u_3=\{v_3,v_6\}$, $u_4=\{v_4,v_5\}$ and $u_\infty=\{v_7,\dots,v_{14}\}$. \label{fig:example2}}
\end{figure}

For example, the BN in Figure \ref{fig:bnex} implies that $X_3$ is conditionally independent of $X_2$ given $X_1$, $p(x_3|x_2,x_1)=p(x_3|x_1)$ for all $x_i\in\mathbb{X}_i$, $i=1,2,3$. The same conditional independence is embedded in the staged tree in Figure \ref{fig:stagedtree} by the staging $\{v_3,v_4\}$ and $\{v_5,v_6\}$ so that $\bm\theta_{v_3}=\bm\theta_{v_4}$ and $\bm{\theta}_{v_5}=\bm\theta_{v_6}$. By construction, all BNs have a staged tree representation such that situations in the same stage must be in the same stratum as in Figure \ref{fig:example}. Only staged trees with this property are implemented in the \pkg{stagedtrees} package. 
\begin{definition}
An $\bm{X}$-compatible staged tree is called stratified if all non-leaf vertices in the same stage are in the same stratum.
\end{definition}
The class of stratified staged trees is much larger than the one of BNs over the same variables: for instance, the staged tree with staging $\{v_3,v_6\}$ and $\{v_4,v_5\}$ in Figure \ref{fig:stagedtree_2} does not have a BN representation over the same X variables. In stratified staged trees the root vertex forms a stage by its own.

Staged trees are very expressive and flexible but, as the number of variables increases, they cannot succinctly visualize their staging. For this reason, \citet{Smith2008} devised a coalescence of the tree by merging some of its vertices in the same stage and therefore reducing the size of the graphical representation. The resulting graph is called a CEG, which represents the exact same probability model as the original staged tree \citep{Collazo2018}. The construction of a CEG from a staged tree is illustrated next.

Given a probability tree $\mathcal{T}$, a subtree $\mathcal{T}(v)$ rooted at $v\in V$ is the tree with $v$-to-leaf paths of $\mathcal{T}$ and the same edge probabilities.  Two vertices $v, v'$ in the same stage are said to be in the same position if the subtrees $\mathcal{T}(v)$ and $\mathcal{T}(v')$ are equal. For instance, the vertices $v_3$ and $v_4$ in Figure \ref{fig:stagedtree} are in the same stage but also in the same position. Therefore, for vertices in the same position the full downstream stage structure is identical, and not only the immediate floret probabilities.  Positions give a coarser partition $U$ of the vertex set of a staged tree than stages do. Hereby, all leaves are trivially in the same position denoted by $u_\infty$. 

The CEG is the graph obtained from a staged tree $\mathcal{T}=(V,E)$ having a vertex for each set in $U$ and edge set $F$ so constructed: if there exist edges $e=(v,v')$, $e'=(w,w')\in E$ and $v, w$ are in the same position then there exist corresponding edges $f,f'\in F$. If also $v',w'$ are in the same position then the labels associated to $f$ and $f'$ are equal and are probabilities inherited from $\mathcal{T}$. The process of constructing a CEG is illustrated in Figure \ref{fig:example}.

\section[Package implementation]{Package implementation} \label{sec:package}

\subsection[Creating staged trees and CEGs]{Creating staged trees and CEGs}

The main object class implemented in the \pkg{stagetrees} package is \code{sevt} representing a staged tree model. Given a dataset, either in \code{data.frame}, \code{table} or \code{list} format, a staged tree which is  compatible with the variables in the dataset can be constructed using the functions \code{full} or \code{indep}. The function \code{full} returns a \code{sevt} object which defines in \proglang{R} a staged tree where each vertex is in a different stage. It corresponds to the saturated statistical model, where the number of free parameters equals the number of edges minus the number of non leaf vertices, equivalently the number of leaves minus one. Conversely, \code{indep} returns a tree where all vertices in the same stratum are in the same stage, corresponding to a model where all variables are marginally independent of each other.

Worth-mentioning arguments of these two functions are: \code{order}, which fixes the order of the variables in the tree; \code{join_unobserved} which collapses parts of the tree where no observations are collected (by default set to \code{TRUE}); \code{lambda}, which implements a Laplace smoothing \citep{Russell2016} to address possible zero counts in case \code{join_unobserved} is set to \code{FALSE}.

Furthermore, a \code{bn.fit} object created with the \pkg{bnlearn} package could be turned into a \code{sevt} object modelling the same conditional independences with  \code{as_sevt}. A staged tree can be converted into a CEG model using the \code{ceg} function. The usual \code{print}, \code{summary} and \code{plot} functions provide basic information, more detailed information and the graphical representation of the model, respectively.

\subsection[Structure learning algorithms]{Structure learning algorithms}
\pkg{stagedtrees} implements a variety of structure learning algorithms. These can be grouped into two categories:
\begin{itemize}
\item score-based algorithms using various heuristics to 
maximize a score function.  The default value of \code{score} is the negative BIC, but any other can be defined by the user:
\begin{itemize}
\item an hill-climbing score optimization implemented in \code{stages_hc} which, for each stratum, at each iteration searches for the vertex to move either to a different or a new stage maximizing a score until no score improvement is found;
\item a backward hill-climbing \code{stages_bhc} which searches the joining of two stages maximizing a score until no score improvement is found;
\item a fast backward hill-climbing \code{stages_fbch} which joins two stages whenever the joining improves the score until no improvement is possible;
\item a random backward hill-climbing \code{stages_bhcr} which at each iteration randomly selects a stratum and two stages and joins the stages if the score is increased. The procedure is repeated until the number of iterations reaches \code{max_iter}.
\end{itemize}
\item Clustering-based algorithms, where stages are created by clustering the probability distribution of florets:
     \begin{itemize}
      \item backward joining of stages \code{stages_bj} which iteratively joins 
      stages if the distance between their floret probabilities is 
      less then a given threshold value (\code{thr}) (the distance can be chosen with the \code{distance} argument, the default being the symmetrized Kullback-Leibler \code{"kullback"});
      \item hierarchical clustering of stages \code{stages_hclust} which creates a user-defined number \code{k} of stages in each stratum. The function inherits all arguments of the standard \code{hclust} function from the \pkg{stats} package;
      \item clustering of stages using the k-means algorithm \code{stages_kmeans}, again creating a user-defined number \code{k} of stages in each stratum. The function inherits all arguments of the standard \code{kmeans} function from the \pkg{stats} package.
     \end{itemize}
\end{itemize}

The starting model of any structure learning algorithm has to be a staged tree which, for instance, may be constructed directly from a dataset using \code{full} or \code{indep}.
Different structure learning algorithm can be easily combined since the starting model for any algorithm could be also an already estimated model with another structure learning algorithm. Furthermore, model search can be computed over a subset of strata specified by \code{scope}.

\subsection[Querying the model]{Querying the model}

\pkg{stagedtrees} provides an array of functions to explore and perform inference over a learned model:
\begin{itemize}
\item \code{stndnaming} standardly renames stages. It assigns them increasing numbers from 1 to the number of different stages, for each stratum in the tree;
\item \code{subtree} enables for the construction of a subtree having as root any vertex of the tree. This can be achieved specifying the \code{path} starting from the root and ending at that vertex;
\item \code{summary} returns for each stratum all the estimated stages, the number of paths and observations starting from the root that arrives to each stage and their corresponding probability distributions;
\item \code{compare_stages} checks if the staging structure of two staged trees with the same order of variables are equal and returns a plot where nodes in different stages are colored in red. 
\item \code{sample_from} generates observations according to the probability distributions defined by the staged tree given in input. This can be used to perform simulation studies over a learned model;
\item \code{get_stage} retrieves the stage associated to a given \code{path} from the root. To be used in combination with \code{summary} and/or \code{plot} for a more helpful use; 
\item \code{get_path} gives all the paths that starting from the root arrive to a given stratum (\code{var}) and \code{stage};
\item \code{prob} computes the probability (or its logarithm if \code{log} = TRUE) of any event of interest (\code{x}) and can be used to derive all atomic probabilities;
\end{itemize}

\subsection{Plotting}

\pkg{stagedtrees} contains simple plotting functions to enable 
a visual exploration and visualization of the generated models. 

\begin{itemize}
\item \code{plot} is a dependencies-free plotting function for staged trees; users can specify stage-colouring, node and edge size and labels appearance. 
\item \code{barplot} automatically generates barplots to visualize the floret probabilities for each stage of a specified variable (\code{var}). 
\end{itemize}

\section[Usage of the stagedtrees package]{Usage of the stagedtrees package} \label{sec:example}


The well-known Titanic dataset \citep{Dawson1995}, which provides information on the fate of the Titanic passengers and available from the \pkg{datasets} package bundled in \proglang{R}, is used to exemplify the usage of \pkg{stagedtrees}. 
\pkg{stagedtrees} and its dependencies (the \pkg{graphics} and \pkg{stats} packages bundled in R) are available from CRAN, as is the package \pkg{bnlearn} \citep{Scutari2010}.

\subsection[Learning the stage structure from a dataset]{Learning the stage structure from a dataset}
\begin{figure}[h!]
\begin{minipage}{0.49\textwidth}
\begin{center}
\includegraphics{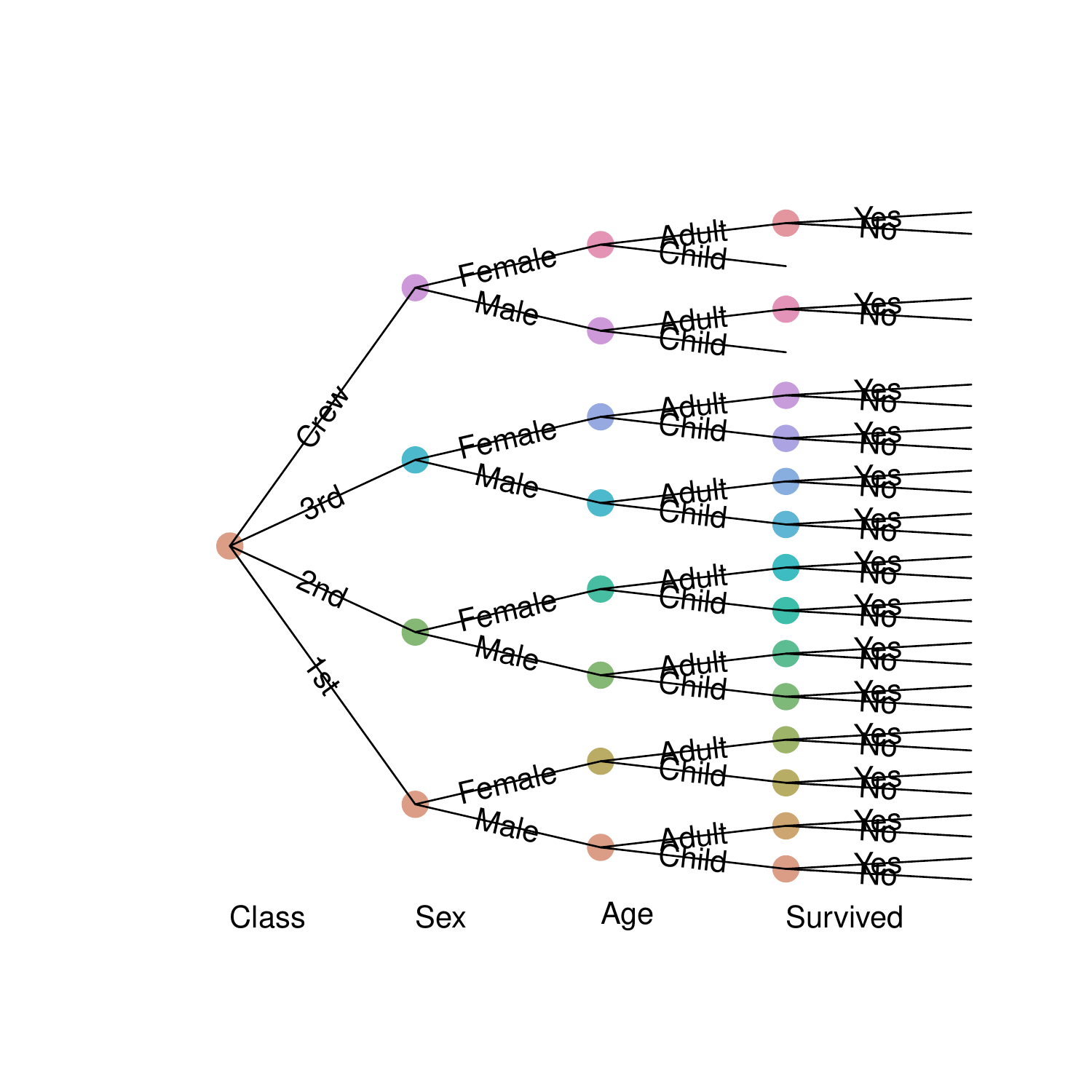} 
\end{center}
\end{minipage}
\begin{minipage}{0.49\textwidth}
\begin{center}
\includegraphics{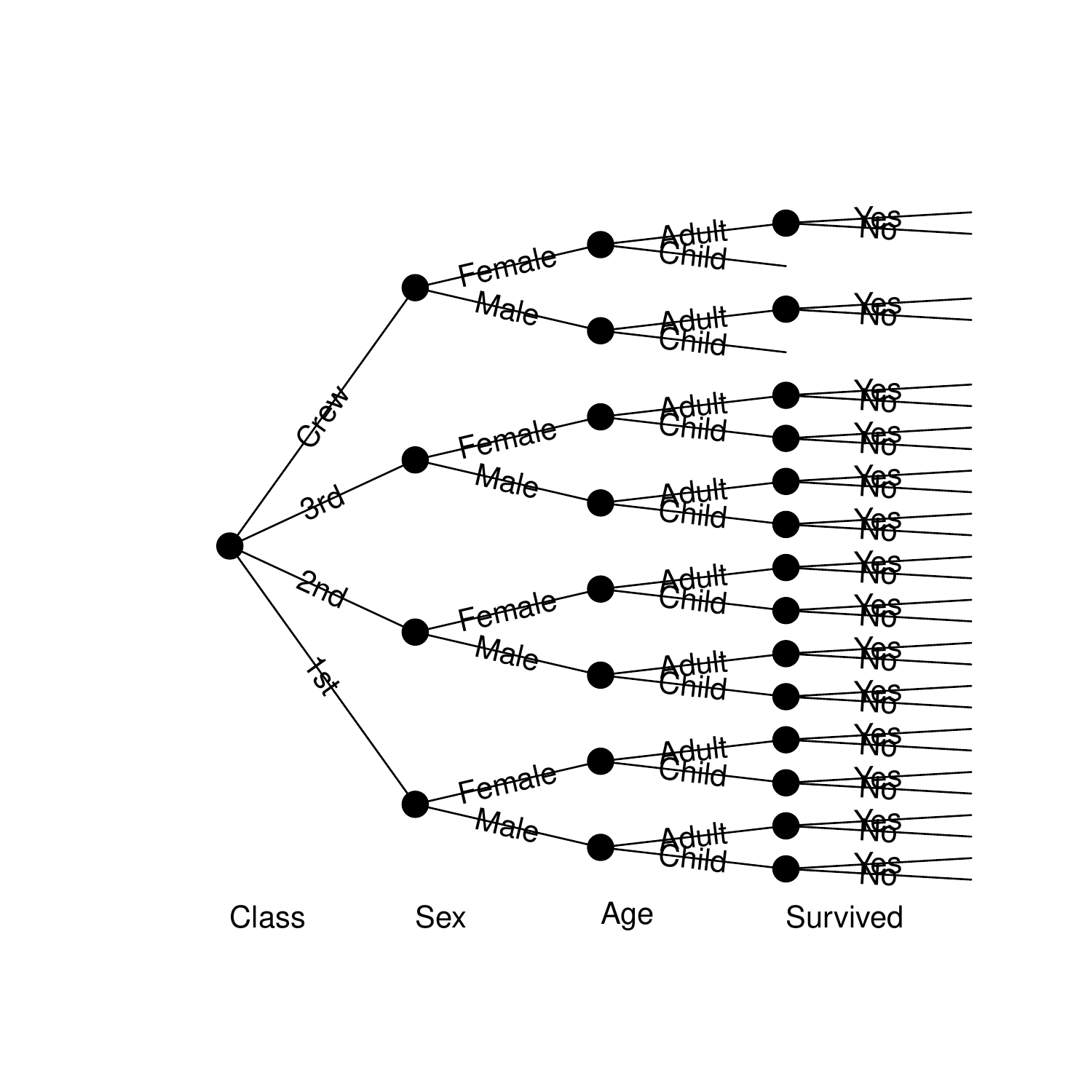} 
\end{center}
\end{minipage}

\caption{\label{fig:ind} Left: Staged tree \code{m.full} where all vertices in the same stratum are in a different stage: there are 29 different stages. Colors in different strata can be equal. Right: Staged tree \code{m.indep} where all vertices in the same stratum are in the same stage: there are 4 different stages. The labels at the bottom denote the variable associated to a stratum.
}
\end{figure}

The \code{Titanic} dataset can be loaded into a \code{table} of the same name with the call to \code{data}.
\begin{Schunk}
\begin{Sinput}
R> data("Titanic")
R> str(Titanic)
\end{Sinput}
\begin{Soutput}
 'table' num [1:4, 1:2, 1:2, 1:2] 0 0 35 0 0 0 17 0 118 154 ...
 - attr(*, "dimnames")=List of 4
  ..$ Class   : chr [1:4] "1st" "2nd" "3rd" "Crew"
  ..$ Sex     : chr [1:2] "Male" "Female"
  ..$ Age     : chr [1:2] "Child" "Adult"
  ..$ Survived: chr [1:2] "No" "Yes"
\end{Soutput}
\end{Schunk}
\code{Titanic} includes four categorical variables: \code{Sex}, \code{Age} and  \code{Survived} are binary and \code{Class} has four levels. Initial staged trees where all vertices within a stratum are either in the same or in different stages can be constructed using the \code{indep} and \code{full} functions, respectively. 
%
\begin{Schunk}
\begin{Sinput}
R> library(stagedtrees)
R> m.full <- full(Titanic, name_unobserved = "na")
R> m.indep <- indep(Titanic, name_unobserved = "na")
R> m.full 
\end{Sinput}
\begin{Soutput}
Staged event tree (fitted) 
Class[4] -> Sex[2] -> Age[2] -> Survived[2]  
'log Lik.' -5151.517 (df=30)
\end{Soutput}
\begin{Sinput}
R> m.indep
\end{Sinput}
\begin{Soutput}
Staged event tree (fitted) 
Class[4] -> Sex[2] -> Age[2] -> Survived[2]  
'log Lik.' -5773.349 (df=7)
\end{Soutput}
\end{Schunk}
The printing of \code{m.full} and \code{m.indep} gives information about the order of the variables in the tree, the value of the log-likelihood function and the number of free parameters, whilst \code{plot} displays the stratified staged tree with stages coloured within each stratum as shown in Figure \ref{fig:ind}. The plot of \code{m.full} is depicted using the \code{Dynamic} palette from the \pkg{colorspace} package \citep{colorspace}, since the default palette has only $8$ colours and thus stages for the last variable would be impossible to graphically distinguish. 

%
\begin{Schunk}
\begin{Sinput}
R> library(colorspace)
R> plot(m.full, col = function(s) qualitative_hcl(length(s), "Dynamic"))
\end{Sinput}
\end{Schunk}
\begin{Schunk}
\begin{Sinput}
R> plot(m.indep)
\end{Sinput}
\end{Schunk}

Notice that there are no crew members, either male or female, who are children and this is correctly reflected in the trees in Figure \ref{fig:ind} since the subtree associated to such events are collapsed (by default the argument \code{join_unobserved} is set to \code{TRUE}). The name of these collapsed vertices is set to \code{"na"} with the argument \code{name_unobserved}.

Using the staged tree \code{m.full} or \code{m.indep} as starting point, structural learning algorithms can be used to infer the staging structure from the data. The hill-climbing algorithm implemented in \code{stages_hc} can receive in input both \code{m.full} and \code{m.indep} (since it embeds also a splitting stage move). Whilst backward algorithms  (implemented in \code{stages_bhc}, \code{stages_fbhc} and \code{stages_bhcr}) and clustering algorithms (implemented in \code{stages_bj}, \code{stages_hclust} and \code{stages_kmeans}) start from the \code{m.full} tree. For illustration purposes, the \code{stages_hc} function is used with the \code{m.indep} tree, whilst \code{stages_bj} is used with \code{m.full}.
%
\begin{Schunk}
\begin{Sinput}
R> mod1 <- stages_hc(m.indep)
R> mod2 <- stages_bj(m.full, thr = 0.1)
\end{Sinput}
\end{Schunk}

\begin{figure}
\begin{minipage}{0.49\textwidth}
\begin{center} 
\includegraphics[width=2.8in]{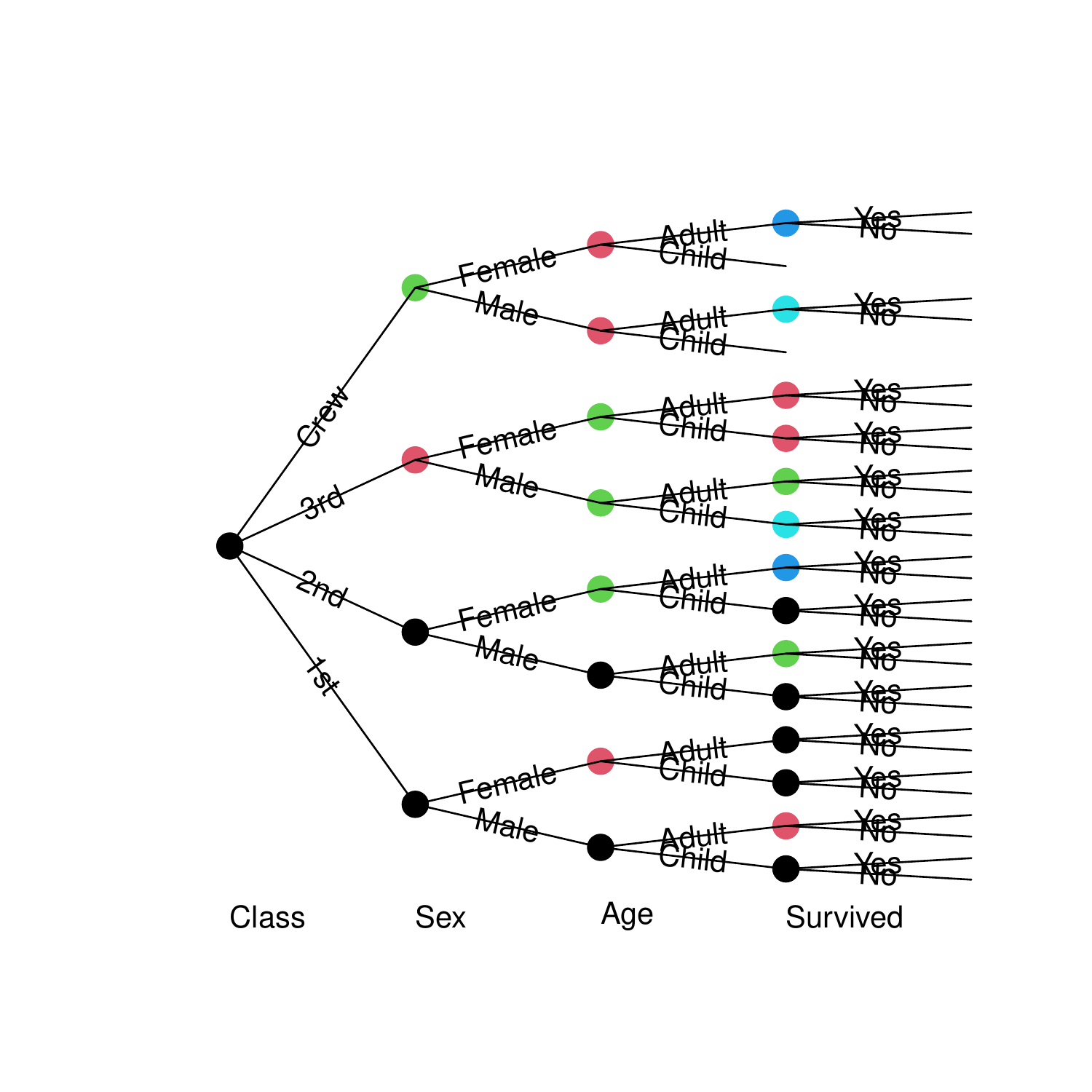} 
\end{center}
\end{minipage}
\begin{minipage}{0.49\textwidth}
\begin{center}
\includegraphics[width=2.8in]{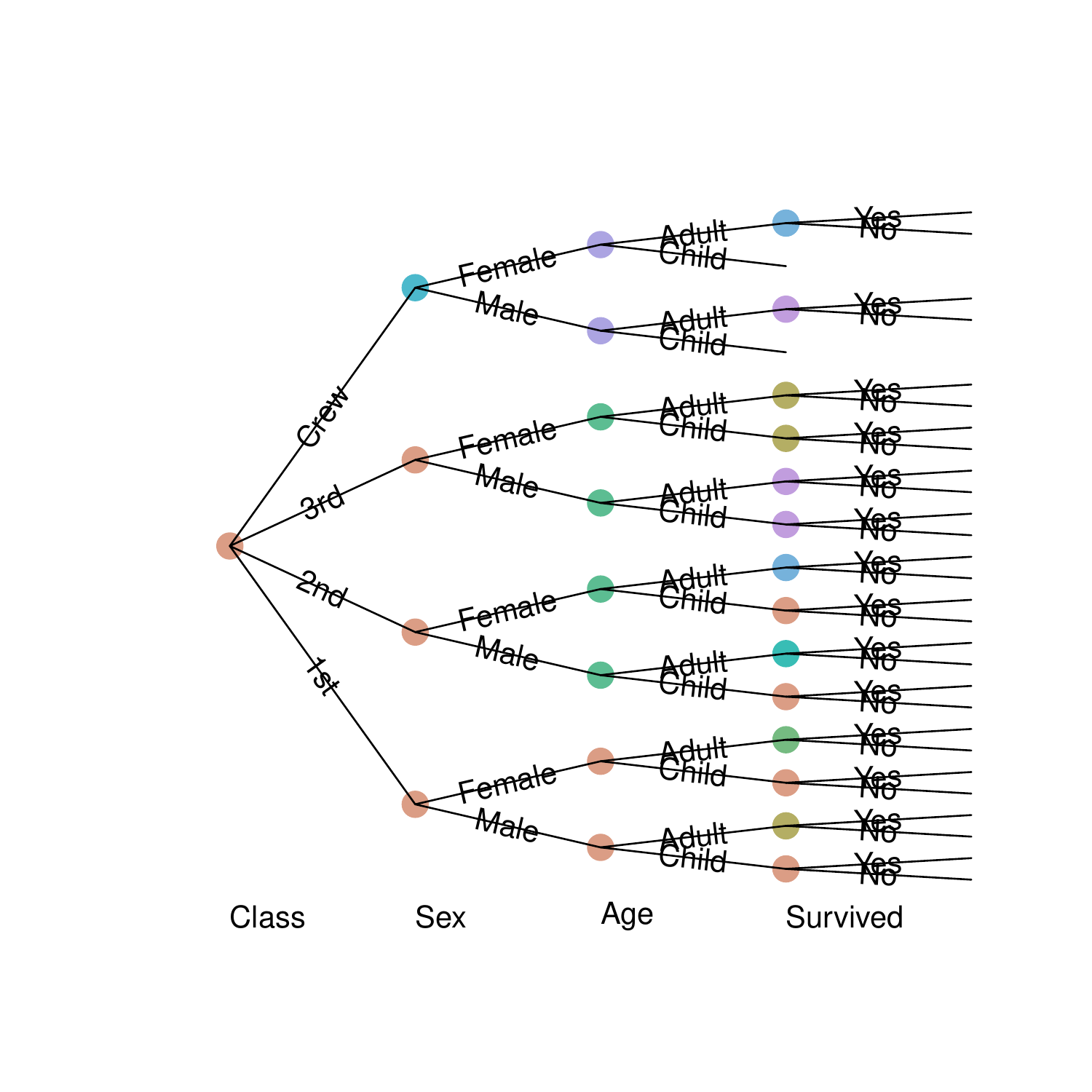} 
\end{center}
\end{minipage}
\caption{\label{fig:learn} Staged trees \code{mod1} (left) and \code{mod2} (right) learned using the \code{stages\_hc} and the \code{stages\_bj} algorithms, respectively.
}
\end{figure}

The \code{stages_hc} function has BIC as a default score, while the default distance for \code{stages_bj} is the symmetrized Kullback-Leibler divergence, with threshold $0.1$ in this example. The learned \code{mod1} and \code{mod2} are plotted in Figure \ref{fig:learn}. Both staged trees suggest that the variables are dependent in a non-symmetric fashion and thus suggest context-specific independences. The stage structures of the two trees are quite different and may be affected by the choice of threshold in \code{mod2}. However, they also share some common features: for instance, both state that the distribution of Male/Female is the same for passengers 
in the first and second class.

\begin{figure}
\begin{minipage}{0.49\textwidth}
\begin{center}
\includegraphics[width=2.8in]{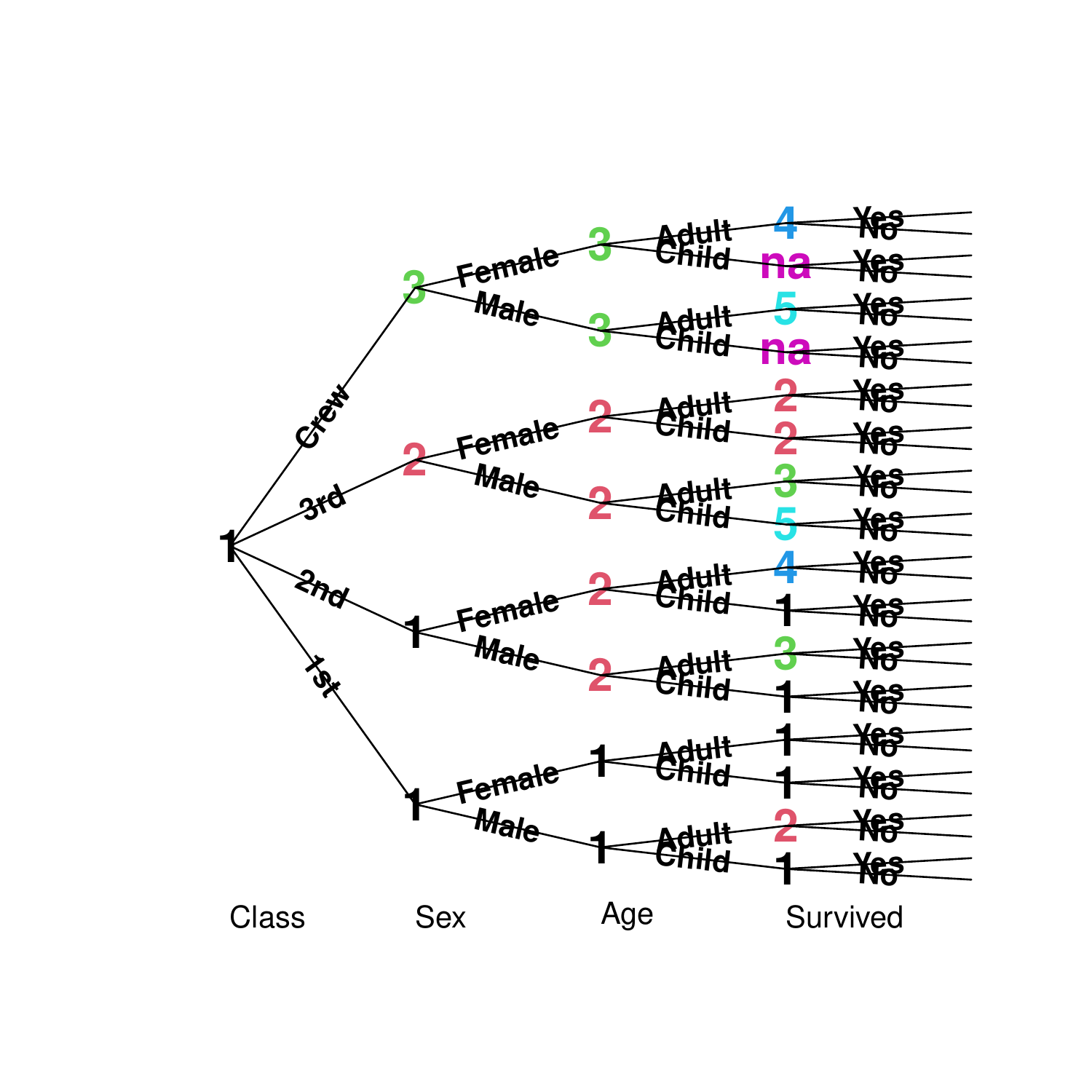} 
\end{center}
\end{minipage}
\begin{minipage}{0.49\textwidth}
\begin{center}
\includegraphics[width=2.8in]{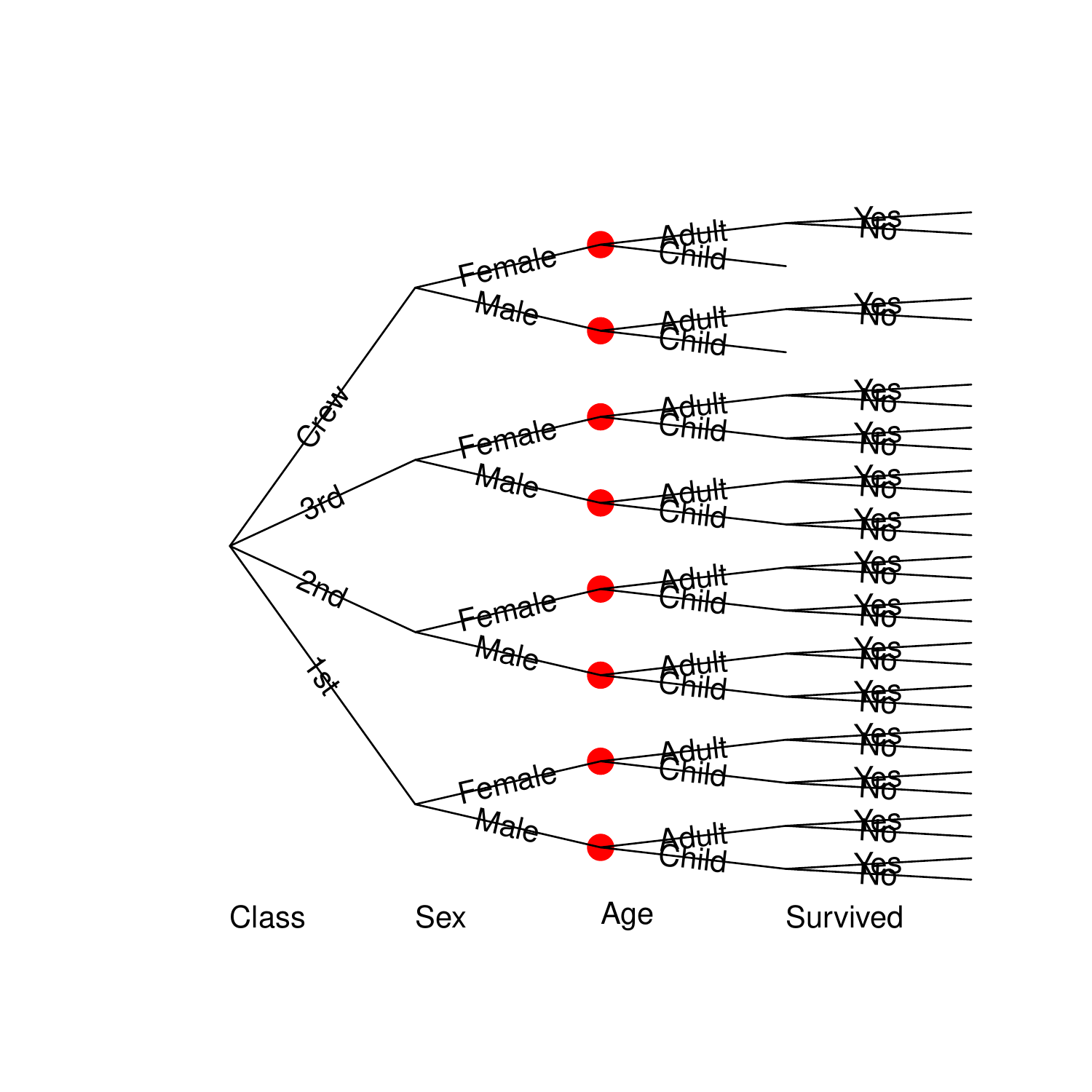} 
\end{center}
\end{minipage}
\caption{\label{fig:model3} Staged event tree \code{mod3} (left) and output of the \code{compare\_stages} function between models \code{mod1} and \code{mod3} (right). Vertices depicted by a red dot in the right plot correspond to vertices for which the staging structure differs.}
\end{figure}

Since all structural learning algorithms take as input a staged tree, it is possible to refine a learned model: for instance the model 
\code{mod2} learned using a backward algorithm may be refined using a standard hill climbing algorithm. 
\begin{Schunk}
\begin{Sinput}
R> mod3 <- stndnaming(stages_hc(mod2))
\end{Sinput}
\end{Schunk}
\begin{Schunk}
\begin{Sinput}
R> plot(mod3, ignore = NULL, 
  +      cex_label_nodes = 1.5, cex_nodes = 0, font = 2)
\end{Sinput}
\end{Schunk}
%
The resulting staged tree is reported in Figure \ref{fig:model3}. For illustrative purpose we report there the full tree (by setting \code{ignore = NULL}) and the numbering of the stages after renaming them with the function \code{stndnaming}. The two staged tree structures in \code{mod1} and \code{mod3} are compared through the \code{compare_stages} function, whose output highlights in red the nodes in different stages.
Different methods can be used to compare two staged tree structures, here the \code{"stages"} method is used: it checks if the same exact stages are present in both models. 

%
\begin{Schunk}
\begin{Sinput}
R> compare_stages(mod1, mod3, method = "stages", plot = TRUE)
\end{Sinput}
\begin{Soutput}
[1] FALSE
\end{Soutput}
\end{Schunk}
%

Figure \ref{fig:model3} shows that the two models have the same stage 
structure over the \code{Sex} and \code{Survived} variables, but they highly differ over 
\code{Age}.

The model selection criteria \code{AIC} and \code{BIC} can be used to choose the best fitting model.
\begin{Schunk}
\begin{Sinput}
R> cbind(AIC(mod1, mod2, mod3), BIC = BIC(mod1, mod2, mod3)$BIC)
\end{Sinput}
\begin{Soutput}
     df      AIC      BIC
mod1 15 10364.49 10449.94
mod2 15 10390.37 10475.82
mod3 15 10365.02 10450.47
\end{Soutput}
\end{Schunk}
According to both criteria, \code{mod1} 
is the best fitting model among those tried. 
It is not surprising that \code{mod2} obtains the worst BIC scores since 
it was estimated with the \code{stages_bj} function that joins stages 
following a distance based heuristic and thus not the minimization of the 
BIC score. 

\subsection{Bayesian networks as staged trees}

\pkg{stagedtrees} has the capability of translating a BN learned 
with the \pkg{bnlearn} package into a staged tree. 
To use \pkg{bnlearn} the dataset \code{Titanic} needs to be 
converted into a data frame.

\begin{figure}[!h]
\begin{minipage}{0.49\textwidth}
\begin{center}
\includegraphics[width=2.8in]{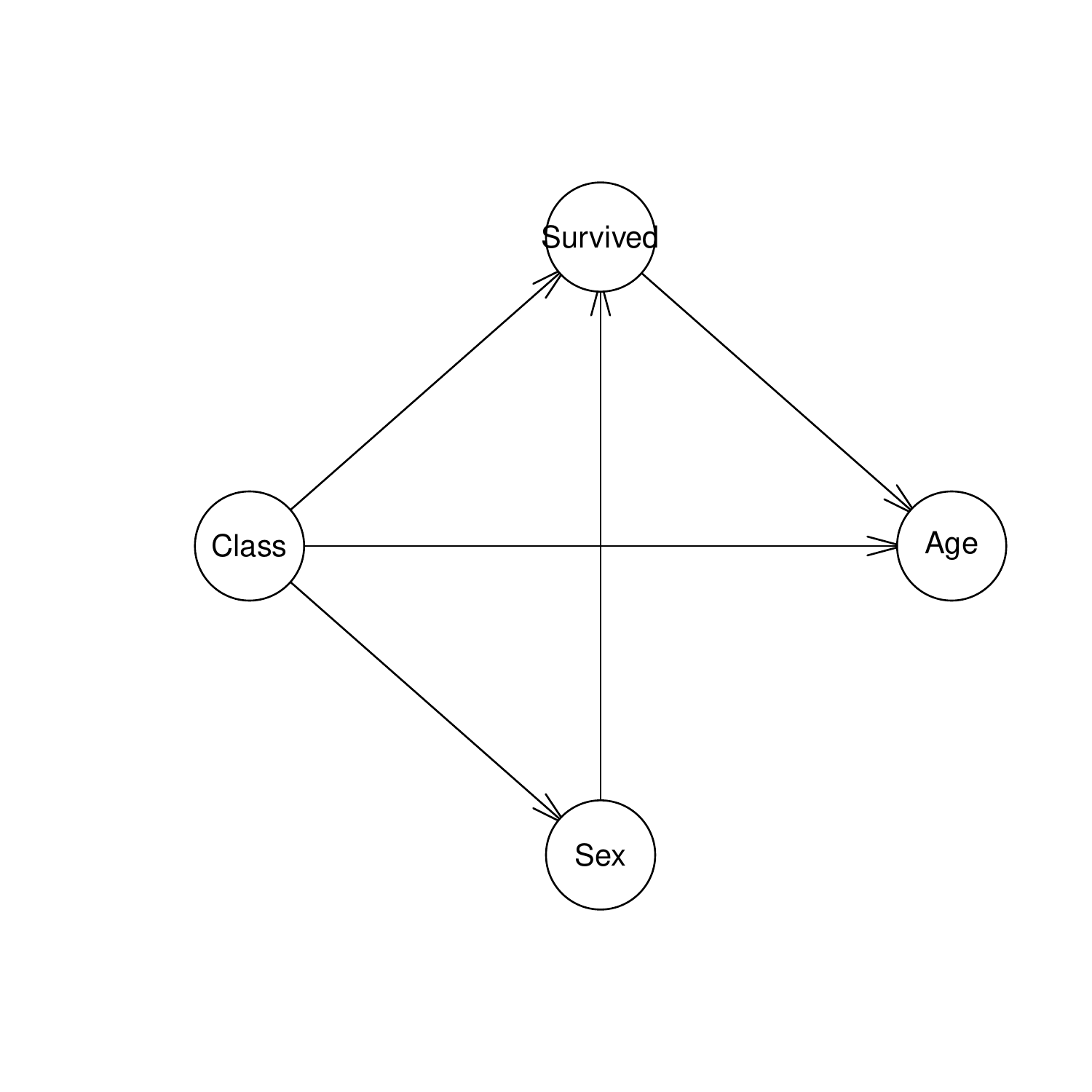} 
\end{center}
\end{minipage}
\begin{minipage}{0.49\textwidth}
\begin{center}
\includegraphics[width=2.8in]{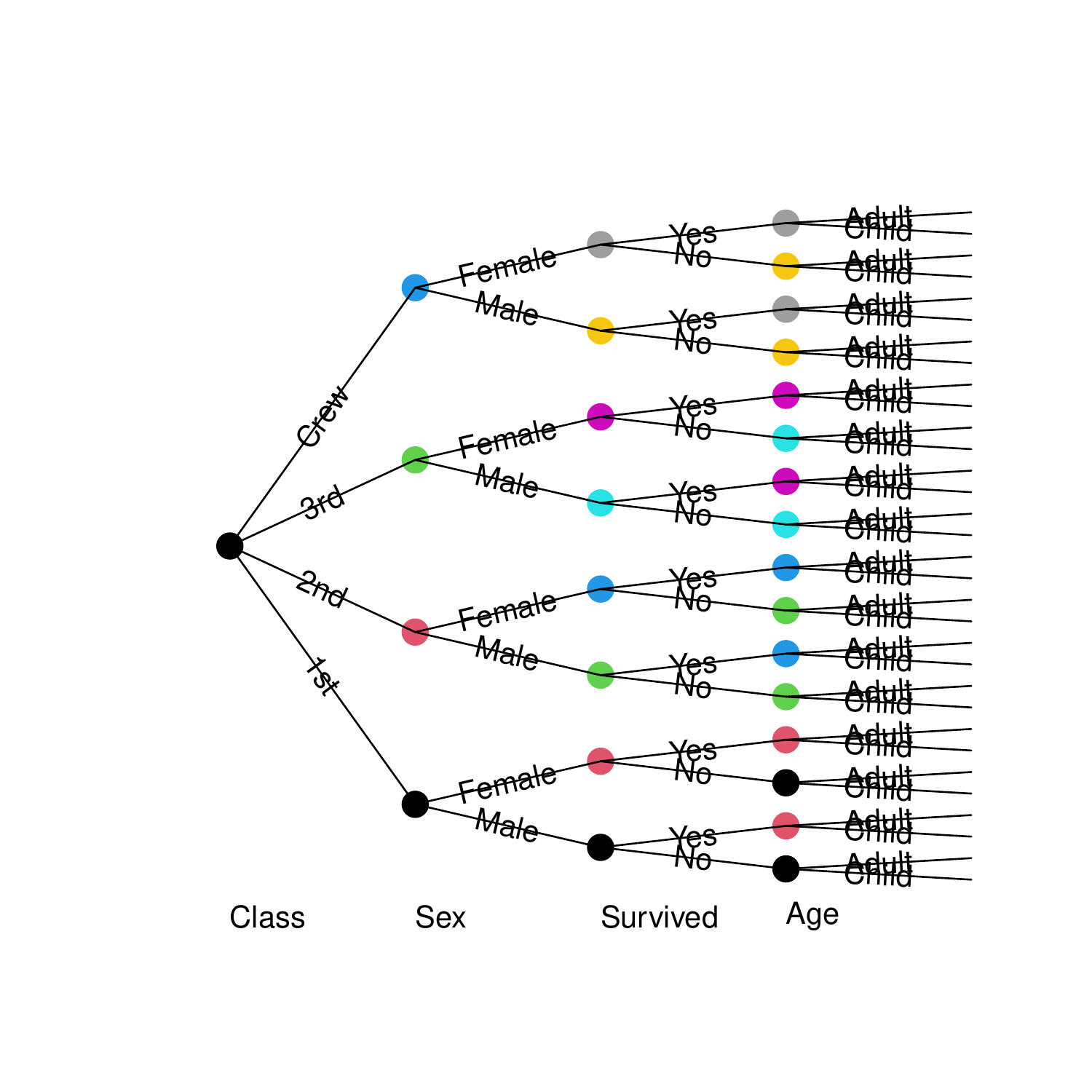} 
\end{center}
\end{minipage}
\caption{\label{fig:bn} Left: BN model learned using the \code{hc} function of \pkg{bnlearn}. Right: associated staged event tree.}
\end{figure}

\begin{Schunk}
\begin{Sinput}
R> titanic.df <- as.data.frame(Titanic)
R> titanic.df <- titanic.df[rep(row.names(titanic.df), titanic.df$Freq), 1:4]
\end{Sinput}
\end{Schunk}
The \code{hc} function of \pkg{bnlearn} can be used to learn the graph of the BN reported in Figure~\ref{fig:bn} left.
\begin{Schunk}
\begin{Sinput}
R> library(bnlearn)
R> mod.bn <- bnlearn::hc(titanic.df)
R> plot(mod.bn)
\end{Sinput}
\end{Schunk}
\code{bn.fit} returns an object of class \code{bn.fit} which can be turned into an object of class \code{sevt} using the \code{as_sevt} function. 
\code{sevt_fit} is used to compute also the stage probability distributions.
Below the \proglang{R} code. 
\begin{Schunk}
\begin{Sinput}
R> mod.bn <- bn.fit(mod.bn, titanic.df)
R> bn.tree <- sevt_fit(as_sevt(mod.bn), data = titanic.df, lambda = 0)
R> plot(bn.tree)
\end{Sinput}
\end{Schunk}

The learned BN embeds only one conditional independence statement: \code{Age} and \code{Sex} are conditionally independent given \code{Class} and \code{Survived}. This is represented in Figure \ref{fig:bn} right by the highly symmetric staging structure over the variable \code{Age}. Notice that this tree, since it is representing the associated BN, does not collapses subtrees where there are no associated observations in the dataset. However this can be achieved by using the function \code{join_unobserved}. It is also worth noticing that the order of the variables chosen by \pkg{bnlearn} is different to the one used for \code{mod1}, \code{mod2} and \code{mod3}. Therefore, it is not possible to use \code{compare_stages} to compare \code{bn.tree} with \code{mod1}, \code{mod2} or \code{mod3}. 

The staged tree corresponding to the associated learned BN could be used as the starting point of any of our structure learning algorithms, as below and also in \citet{Barclay2013}. As an illustration, we use here the \code{stages_hclust} function specifying that in each stratum there should be 2 stages.

\begin{Schunk}
\begin{Sinput}
R> mod4 <- stages_hclust(bn.tree, k = 2)
R> plot(mod4, col = function(x) c("red3", "blue3"))
\end{Sinput}
\end{Schunk}

\begin{figure}[!h]
\begin{minipage}{0.49\textwidth}
\begin{center}
\includegraphics{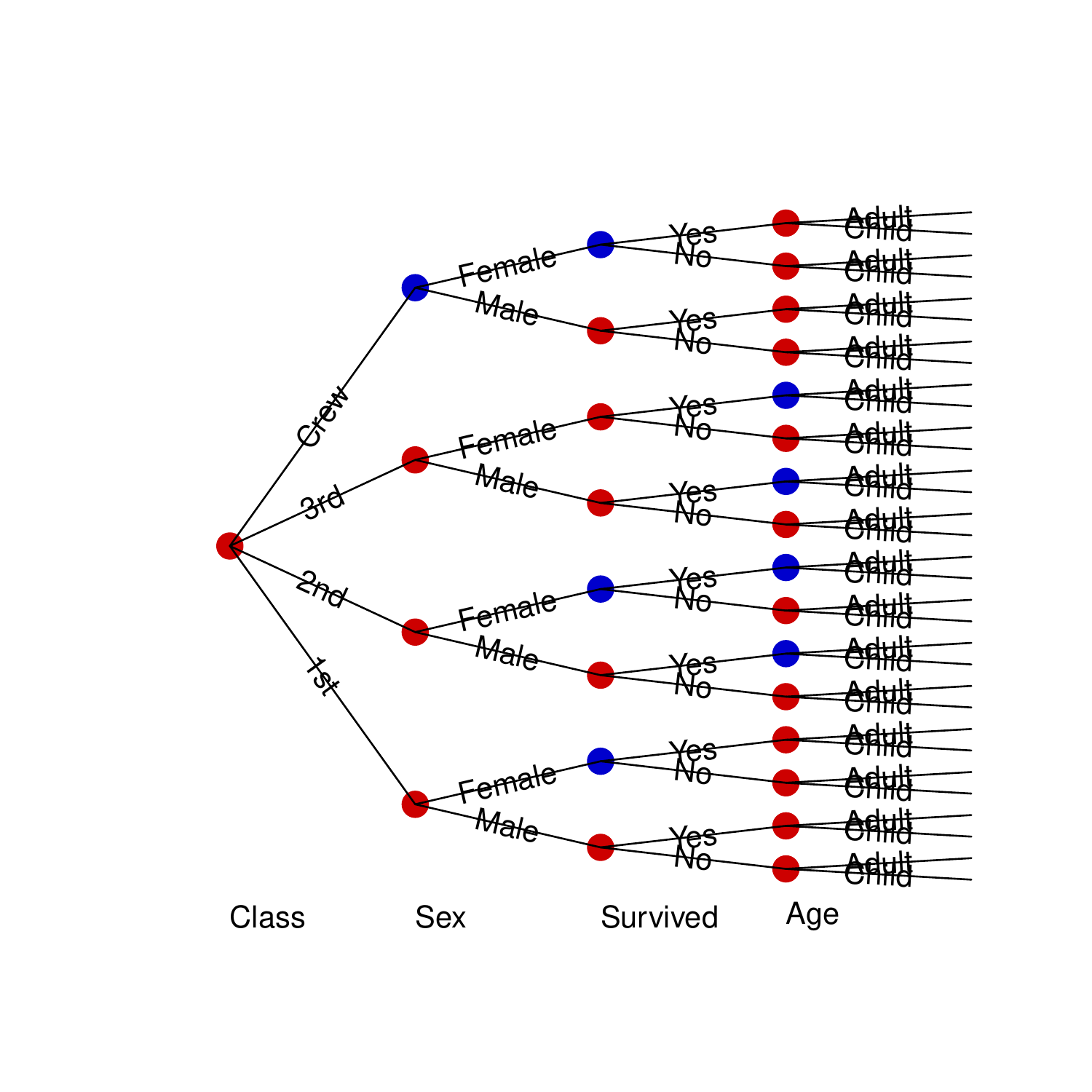} 
\end{center}
\end{minipage}
\begin{minipage}{0.49\textwidth}
\begin{center}
\includegraphics{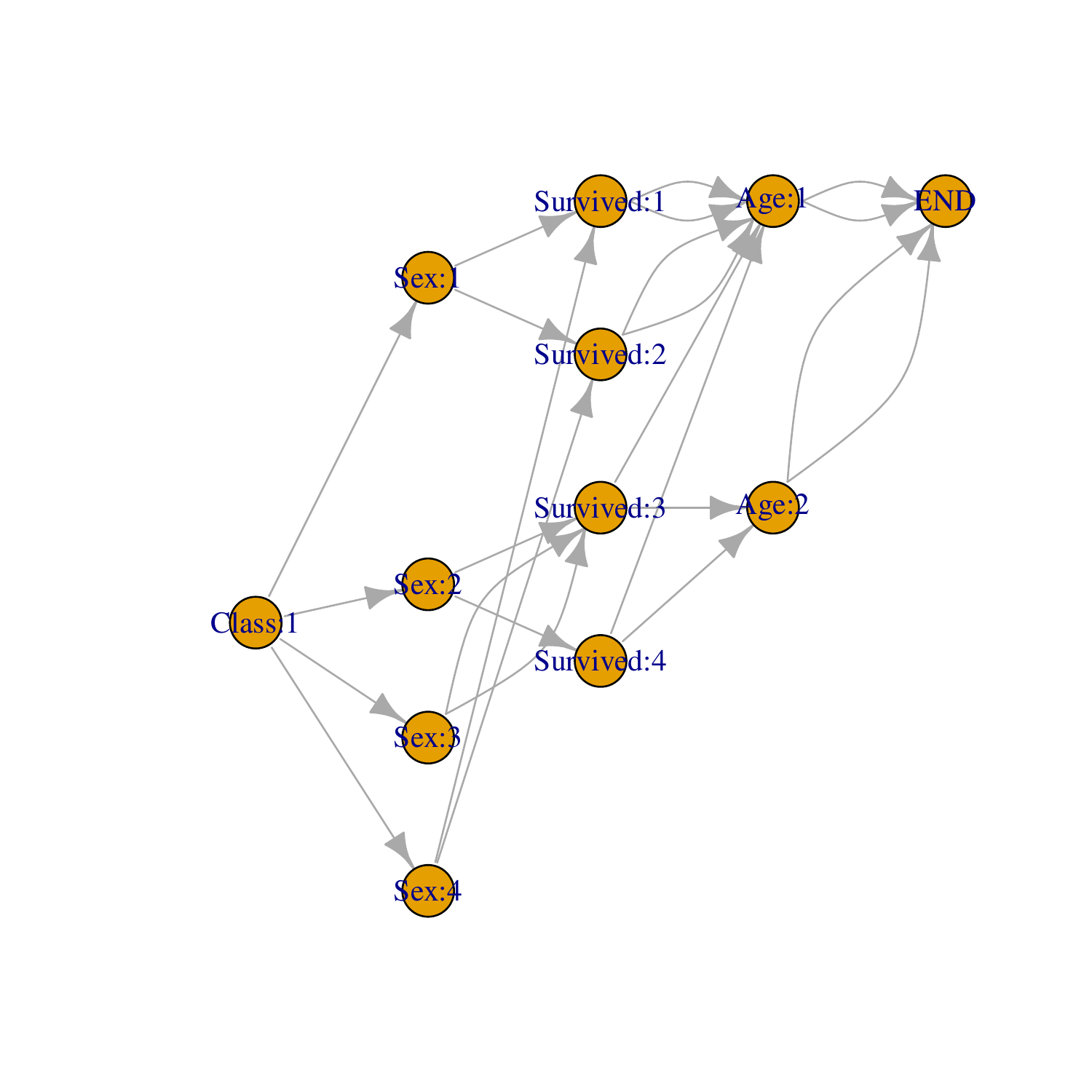} 
\end{center}
\end{minipage}
\caption{\label{fig:ceg} Staged event tree \code{mod4} (left) and its corresponding CEG representation (right).}
\end{figure}

The staged tree \code{mod4}, which is displayed in Figure \ref{fig:ceg} left, is coalesced into the more compact CEG representation. This can be achieved by the \code{ceg} function which takes as input \code{mod4}.
\begin{Schunk}
\begin{Sinput}
R> library(igraph)
R> ceg <- ceg(mod4)
R> A <- ceg2adjmat(ceg)
R> gr <- graph_from_adjacency_matrix(A)
R> lay = layout.reingold.tilford(gr)
R> plot.igraph(gr, layout = -lay[, 2:1])
\end{Sinput}
\end{Schunk}
The resulting CEG plot, which was produced using the \pkg{igraph} package \citep{igraph}, is shown in Figure \ref{fig:ceg} right. For this model, vertices in the last stratum are coalesced into two positions, whilst vertices in the penultimate stratum are coalesced into four positions, thus reducing the overall number of vertices of the underlying graphical representation.


\subsection[Querying the model]{Querying the model}

Chosen a model, the focus is on using it to perform inference and understanding the relationship between the problem variables. Here we choose \code{mod1} which was the best scoring model according to AIC and BIC.

The dataset in this simple example only includes four variables and its staged tree can be easily investigated by eye. For more complex applications the function \code{subtree} is useful as it enables the construction of a subtree having as root any vertex of the tree. This can be achieved specifying the path starting from the root and ending at that vertex. For instance, it is possible to construct the subtree relative to the crew of the Titanic.
\begin{Schunk}
\begin{Sinput}
R> subtree.crew <- subtree(mod1, c(Class = "Crew"))
R> subtree.crew
\end{Sinput}
\begin{Soutput}
Staged event tree (fitted) 
Sex[2] -> Age[2] -> Survived[2]  
\end{Soutput}
\begin{Sinput}
R> plot(subtree.crew)
\end{Sinput}
\end{Schunk}
\begin{figure}[!h]
\begin{center}
\begin{minipage}{0.6\textwidth}
\includegraphics{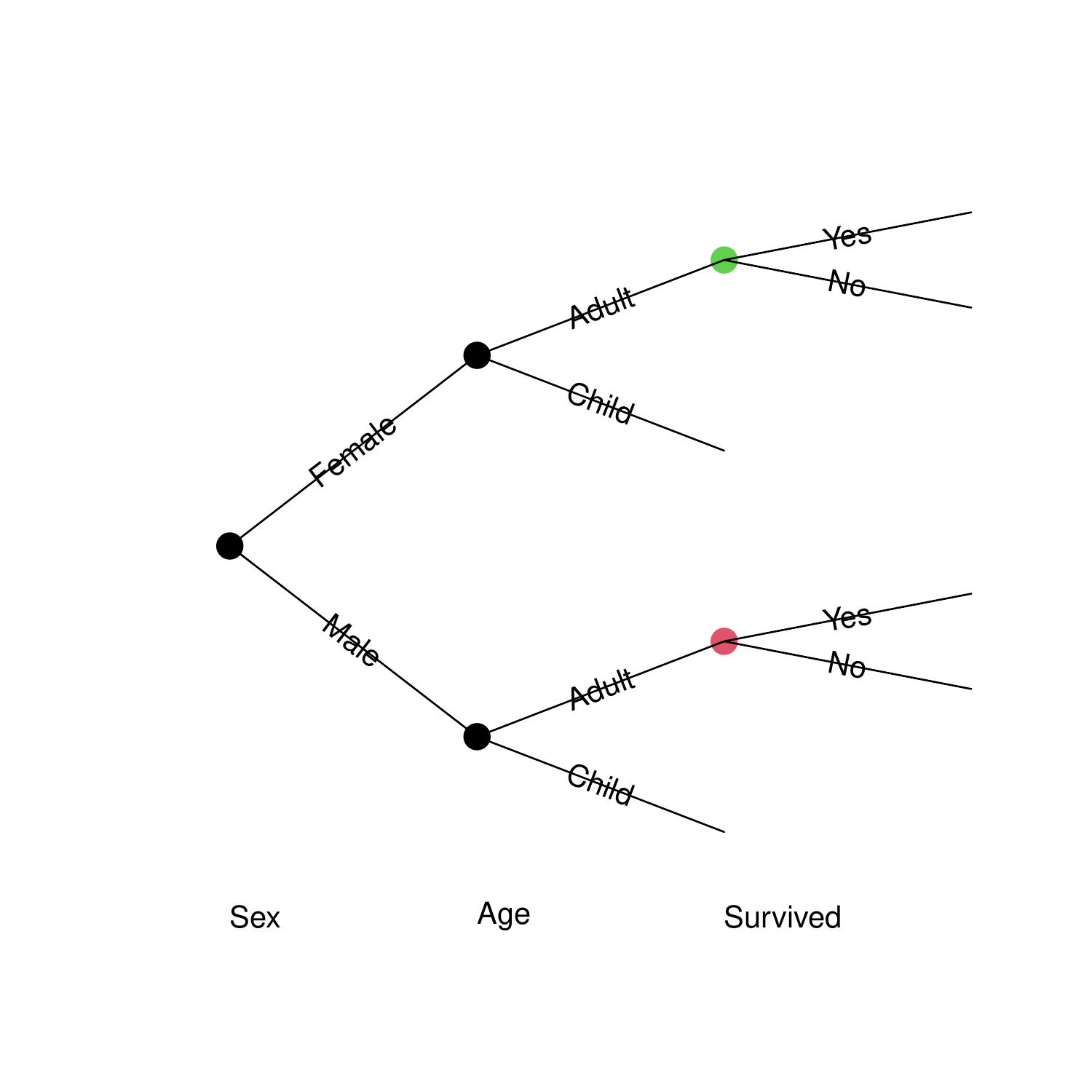} 
\end{minipage}
\end{center}
\vspace{-1cm}
\caption{Subtree of the staged tree \code{mod1} representing \code{Sex}, \code{Age} and \code{Survived} of \code{Crew} passengers only. \label{fig:subtree}}
\end{figure}

\code{subtree.crew} is still formally a staged tree over three variables. The subtree is displayed in Figure \ref{fig:subtree} and its stage structure coincides with the one in the upper half of \code{mod1} reported on the left of Figure \ref{fig:learn}. The colors of the stages are different in the two plots since two different colors palettes have been used.

A detailed model summary of \code{mod1} can be obtained by the 
\code{summary} function. For ease of exposition we also rename the stages with \code{stndnaming}.
\begin{Schunk}
\begin{Sinput}
R> mod1 <- stndnaming(mod1)
R> summary(mod1)
\end{Sinput}
\begin{Soutput}
Call: 
stages_hc(m.indep)
lambda:  0 
Stages: 
  Variable:  Class 
 stage npaths sample.size       1st       2nd       3rd    Crew
     1      0        2201 0.1476602 0.1294866 0.3207633 0.40209
  ------------ 
  Variable:  Sex 
 stage npaths sample.size      Male    Female
     1      2         610 0.5885246 0.4114754
     2      1         706 0.7223796 0.2776204
     3      1         885 0.9740113 0.0259887
  ------------ 
  Variable:  Age 
 stage npaths sample.size        Child     Adult
     1      2         359 0.0445682451 0.9554318
     2      3        1030 0.0009708738 0.9990291
     3      3         812 0.1133004926 0.8866995
  ------------ 
  Variable:  Survived 
 stage npaths sample.size         No       Yes
     1      5         174 0.02298851 0.9770115
     2      3         371 0.60377358 0.3962264
     3      2         630 0.85873016 0.1412698
     4      2         116 0.13793103 0.8620690
     5      2         910 0.77472527 0.2252747
    na      2           0         NA        NA
  ------------ 
\end{Soutput}
\end{Schunk}
The output of \code{summary} together with the function \code{get_stage} allow us to determine the estimated  survival probabilities of the passengers of the Titanic. Stage 1 for \code{Survived} has the highest survival probability and it includes children from the first two classes and adult women from the first class as shown by the following code.
\begin{Schunk}
\begin{Sinput}
R> get_path(mod1, var = "Survived", stage = "1")
\end{Sinput}
\begin{Soutput}
  Class    Sex   Age
1   1st   Male Child
3   1st Female Child
4   1st Female Adult
5   2nd   Male Child
7   2nd Female Child
\end{Soutput}
\end{Schunk}
Stage  3 has the lowest survival probability and includes adult males of second and third class.
\begin{Schunk}
\begin{Sinput}
R> get_path(mod1, var = "Survived", stage = "3")
\end{Sinput}
\begin{Soutput}
   Class  Sex   Age
6    2nd Male Adult
10   3rd Male Adult
\end{Soutput}
\end{Schunk}

The package \pkg{stagedtrees} also includes the function \code{get_stage} to get the stage associated to a given path.
\begin{Schunk}
\begin{Sinput}
R> get_stage(mod1, path = c("Crew", "Female"))
\end{Sinput}
\begin{Soutput}
[1] "2"
\end{Soutput}
\end{Schunk}
The function \code{prob} allows for the computation of the probability of any event of interest.
\begin{Schunk}
\begin{Sinput}
R> prob(mod1, c(Survived = "Yes"))
\end{Sinput}
\begin{Soutput}
[1] 0.3236376
\end{Soutput}
\begin{Sinput}
R> prob(mod1, c(Survived = "Yes", Age = "Adult")) / prob(mod1, c(Age = 
  + "Adult"))
\end{Sinput}
\begin{Soutput}
[1] 0.3165252
\end{Soutput}
\begin{Sinput}
R> prob(mod1, c(Survived = "Yes", Age = "Child")) / prob(mod1, c(Age = 
  + "Child"))
\end{Sinput}
\begin{Soutput}
[1] 0.4584954
\end{Soutput}
\end{Schunk}
For instance, the probability of survival of any passenger is 0.3236, but this decreases to 0.3165 or increases to  0.4585 given that the passenger was an adult or a child, respectively. All atomic probabilities related to the leaves of the staged tree can be also obtained as follows:
\begin{Schunk}
\begin{Sinput}
R> obs <- expand.grid(mod1$tree[4:1])[, 4:1]
R> cbind(obs, p = round(prob(mod1, obs), 6))
\end{Sinput}
\begin{Soutput}
   Class    Sex   Age Survived        p
1    1st   Male Child       No 0.000089
2    1st   Male Child      Yes 0.003784
3    1st   Male Adult       No 0.050130
4    1st   Male Adult      Yes 0.032898
5    1st Female Child       No 0.000001
6    1st Female Child      Yes 0.000058
7    1st Female Adult       No 0.001395
8    1st Female Adult      Yes 0.059304
9    2nd   Male Child       No 0.000078
10   2nd   Male Child      Yes 0.003318
11   2nd   Male Adult       No 0.062524
12   2nd   Male Adult      Yes 0.010286
13   2nd Female Child       No 0.000139
14   2nd Female Child      Yes 0.005898
15   2nd Female Adult       No 0.006516
16   2nd Female Adult      Yes 0.040727
17   3rd   Male Child       No 0.020339
18   3rd   Male Child      Yes 0.005914
19   3rd   Male Adult       No 0.176434
20   3rd   Male Adult      Yes 0.029025
21   3rd Female Child       No 0.006092
22   3rd Female Child      Yes 0.003998
23   3rd Female Adult       No 0.047675
24   3rd Female Adult      Yes 0.031286
25  Crew   Male Child       No 0.000000
26  Crew   Male Child      Yes 0.000000
27  Crew   Male Adult       No 0.303119
28  Crew   Male Adult      Yes 0.088141
29  Crew Female Child       No 0.000000
30  Crew Female Child      Yes 0.000000
31  Crew Female Adult       No 0.001440
32  Crew Female Adult      Yes 0.009000
\end{Soutput}
\end{Schunk}
It shows that around 30\% of the observations follows the root-to-leaf path \code{Crew}, \code{Male}, \code{Adult}, \code{No}.

\begin{figure}[!h]
\begin{center}
\begin{minipage}{0.49\textwidth}
\begin{center}
\includegraphics[scale = 0.2]{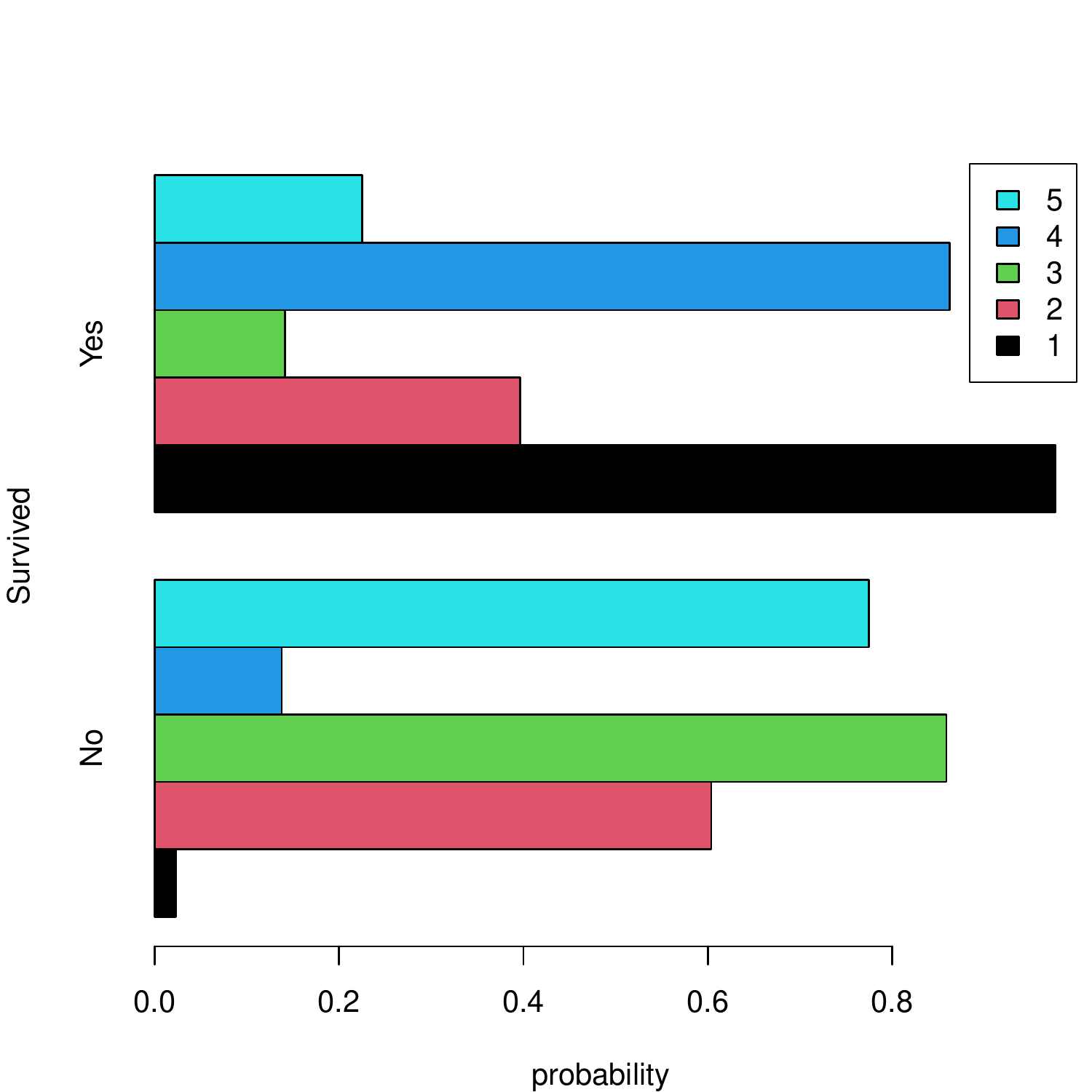} 
\end{center}
\end{minipage}
\end{center}
\caption{\label{fig:model1a} Output of the \code{barplot} function for the variable \code{Survived} according to the stage structure of \code{mod3} depicted in Figure~\ref{fig:model3} left.}
\end{figure}

Finally, barplots can be created to give a visual representation of the estimated probabilities associated to a stratum of the tree as reported in Figure \ref{fig:model1a}.

%
\begin{Schunk}
\begin{Sinput}
R> barplot(mod3, "Survived", legend.text = TRUE, horiz = TRUE, 
  + args.legend = list(x = 1), ylab = "Survived")
\end{Sinput}
\end{Schunk}

\section[A comparison analysis]{A comparison analysis} \label{sec:comparison}
A comparison analysis of structural learning algorithms implemented in \pkg{stagedtrees} is performed on ten datasets, chosen mostly from the literature on CEGs and probabilistic graphical models for contingency tables. The main features of the datasets are summarized in Table~\ref{tab:comparison}, which for each dataset gives the number of observations, variables, root-to-leaf path, cells with zero counts (either observed or structural), non-leaf nodes and edges in the staged tree.
The datasets are available from the \pkg{stagedtrees}, \pkg{datasets} and \pkg{gRbase} \citep{gRbase} \proglang{R} packages. 
It is not the purpose of this section to show how to model these datasets. For this we refer to Section~\ref{sec:interpret} and to the main references for each dataset reported in Table~\ref{tab:ref}.

\begin{table}[!h]
\centering
\small
\scalebox{0.89}{
\begin{tabular}{c|cccccc}
 \hline
 Dataset & \# observations & \# variables & \# root-to-leaf paths $\lambda$ & \# non-leaf nodes & \# 0 cells & \# edges \\ 
  \hline
  \code{Asym} & 1000 & 4 & 16 & 15 & 1 & 30 \\ 
  \code{chestSim500} & 500 & 8 & 256 & 255 & 182 & 510 \\ 
  \code{FallEld} & 50000 & 4 & 64 & 27 & 0 & 90 \\ 
  \code{monks1} & 432 & 7 & 864 & 603 & 243 & 1466 \\ 
  \code{PhDArticles} & 915 & 6 & 144 & 136 & 0 & 279 \\ 
  \code{Pokemon} & 999 & 5 & 32 & 31 & 0 & 62 \\ 
  \code{puffin} & 69 & 6 & 768 & 343 & 284 & 1110 \\ 
  \code{reinis} & 1841 & 6 & 64 & 63 & 0 & 126 \\ 
  \code{selfy} & 2804 & 4 & 72 & 34 & 4 & 105 \\ 
  \code{Titanic} & 2201 & 4 & 32 & 27 & 0 & 58 \\ 
   \hline
\end{tabular}
}
\caption{Summary information about the ten datasets considered for the comparison analysis in Section~\ref{sec:comparison}. \label{tab:comparison}}
\end{table}

\begin{table}[!h]
\centering
\small
\begin{tabular}{c|cc}
  \hline
 Dataset & References & \proglang{R} Package \\ 
  \hline
  \code{Asym} & simulated dataset & \pkg{stagedtrees} \\
  \code{chestSim500} & \cite{hojsgaard2012graphical} & \pkg{gRbase} \\ 
  \code{FallEld} & \cite{10.1007/978-3-030-30611-3_16} &  \\ 
  \code{monks1} & \citet{MONKS} & \\
  \code{PhDArticles} & \cite{Long1990} & \pkg{stagedtrees} \\ 
  \code{Pokemon} & \cite{gabbiadini2018does} & \pkg{stagedtrees} \\ 
  \code{puffin} & \cite{bouveyron2019model} & \pkg{MBCbook} \\
  \code{reinis} & \cite{hojsgaard2012graphical} & \pkg{gRbase} \\ 
  \code{selfy} & \cite{dalla2019catching} & \\ 
  \code{Titanic} & \cite{Dawson1995} & \pkg{datasets} \\ 
   \hline
\end{tabular}
\caption{Main references and \proglang{R} packages related to the analyzed datasets. \label{tab:ref}}
\end{table}

A short simulation study over these ten datasets is conducted. Twelve algorithms from the \pkg{stagedtrees} package are run on each dataset (all score-based algorithms use BIC as \code{score}). Seven additional models from the literature are estimated, namely BNs using hill-climbing and tabu search (in \pkg{bnlearn}), Naive Bayes Classifiers \citep[in \pkg{e1071}][]{e1071}, Logistic Regression and Neural Networks with $10$ units in the hidden layer and weight decay equal to $0.001$ \citep[in \pkg{nnet}][]{nnet}, Classification Trees and Random Forests with $200$ trees and three variables randomly sampled as candidates at each split \citep[in \pkg{rpart} and \pkg{randomForest,}][respectively]{rpart,RandomForest}. See Table~\ref{tab:alg} for details.

For each dataset, each algorithm is run 10 times on 80\% of the data randomly selected and the estimated model is tested on the remaining 20\% of the dataset. The average of all the investigated quantities over the 10 runs is then computed. We compute the number of degrees of freedom, log-likelihood, AIC and BIC values, classification accuracy (the classification variable is the one in the first stratum of the staged tree)  and computational cost of models estimated with 12 algorithms in \pkg{stagedtrees}. 

For ease of exposition, we report here in Table \ref{tab:selfy} the results over the \code{selfy} dataset for the 12 algorithms from \pkg{stagedtree}, although similar conclusions could be drawn from any other dataset. For all datasets we report in Table \ref{table:literature} the mean accuracies of the algorithms from the literature as well as the mean accuracy of the staged tree learnt with \code{stages_bhc}, as a representative from the \pkg{stagedtrees} package. The following general conclusions can be made based on the results reported in these tables: 

\begin{table}
\centering
\begin{tabular}{c|ccc}
\hline
 & Name & Function (\proglang{R} Package) \\
 \hline
\multirow{10}{*}{\pkg{stagedtrees}} & Indep & \code{indep} \\
& Full & \code{full} \\
& HC - Indep & \code{stages\_hc} \\
& HC - Full & \code{stages\_hc} \\
& BHC & \code{stages\_bhc} \\
& Fast BHC & \code{stages\_fbhc} \\
& Random BHC & \code{stages\_bhcr} \\
& Kullback-Leibler & \code{stages\_bj},  \\
& Refined BN & \code{stages_bhc(as_sevt(bn.fit( )))}  \\
& HClust & \code{stages\_hclust} \\
\hline
\multirow{7}{*}{\pkg{literature}} & Bnlearn Hill-Climbing & \code{hc} (\pkg{bnlearn}) \\
& Bnlearn Tabu & \code{tabu} (\pkg{bnlearn}) \\
& Naive Bayes Classifier & \code{naiveBayes} (\pkg{e1071})  \\
& Logistic Model & \code{multinom} (\pkg{nnet}) \\
& Neural Network & \code{nnet} (\pkg{nnet})  \\
& Classification Tree & \code{rpart} (\pkg{rpart})  \\
& Random Forest & \code{randomForest} (\pkg{randomForest}) \\
\hline
\end{tabular}
\caption{List of the algorithms from the \proglang{R} package \pkg{stagedtrees} and from the literature used for model estimation on the ten datasets in Table~\ref{tab:ref}. In round brackets the \proglang{R} packages used. \label{tab:alg}}
\end{table}

\begin{itemize}
\item Full and Indep are the starting models in order to compare the performances of all the structural learning algorithms implemented. The first fits a full-dependence structure to the dataset, by providing one of the best results according to the log-likelihood, due to the over-fitting introduced. The Indep model fits a full-independence structure to the dataset, estimating always the smallest log-likelihood, due to its under-fitting.
\item The number of estimated parameters (df) is highly variable, according to the criterion and the starting stage structure (dependence or independence model). As expected, for backward algorithms with joining based on the Kullback-Leibler distance, the higher is the threshold below which the distance between the transition distributions of two stages are set to be equal, the lower will be the number of estimated parameters.
\item Most often, the higher the number of degrees of freedom of a model has, the higher will be the correspoding log-likelihood value. 
\item The minimum values of the AIC and BIC indices are attained with hill-climbing algorithms. This is intuitive, because the implemented score-based algorithms have as optimization default the minimization of the BIC index. However, even if the distance-based algorithms not aiming to minimize these indices, their performances according to AIC and BIC values are satisfactory and comparable with the score-based methods.
\item The hill-climbing algorithms are slower than others. In particular, the hill-climbing starting from the full-dependence model (HC - Full) is the slowest, because it both joins and splits stages. Conversely,  distance-based methods, fast or random backward hill-climbing and HClust are the fastest. 
\item The accuracy of all models is comparable, the lowest scoring models being Indep and HClust due to their simplicity.
\item  The accuracy of the \pkg{stagedtrees} BHC algorithm is higher in almost all datasets than the one of Bayesian network models, thus highlighting the need for context-specific conditional independence models in real-world applications.
\item The simulated \code{Asym} dataset is characterised by context-specific conditional independences.  As expected from the theory, all proposed algorithms in stagedtrees give better accuracies than ones obtained with Bayesian networks.
\item Overall the algorithms implemented in \pkg{stagedtrees} have competitive accuracy, although these structural learning algorithms have the aim to estimate the joint probability distribution and not the conditional one of interest as for most of the literature algorithms. More precisely, all the literature's models in Table~\ref{tab:alg}, except the ones from \pkg{bnlearn}, estimate directly the conditional probability of observing the response variable, given all the other explanatory variables.
\end{itemize}

\begin{table}[!h]
\centering
\scalebox{0.92}{
\begin{tabular}{ccccccc}
  \hline
 Algorithm & df & logLik & AIC & BIC & Accuracy & Computational Time \\ 
  \hline
Indep & 10.00 & -7892.51 & 15805.03 & 15862.19 & 0.7554 & 0.2396 \\ 
  Full & 64.60 & -6251.36 & 12631.92 & 13001.18 & 0.8495 & 0.2459 \\ 
  HC - Indep & 31.00 & -6277.94 & 12617.89 & 12795.08 & 0.8489 & 1.1130 \\ 
  HC - Full & 35.20 & -6264.32 & 12599.04 & 12800.25 & 0.8507 & 3.5747 \\ 
  BHC & 32.60 & -6271.70 & 12608.60 & 12794.94 & 0.8491 & 0.3916 \\ 
  Fast BHC & 31.00 & -6301.09 & 12664.18 & 12841.38 & 0.8495 & 0.2466 \\ 
  Random BHC & 37.60 & -6284.50 & 12644.21 & 12859.13 & 0.8480 & 0.2476 \\ 
  Kullback-Leibler - 0.01 & 60.40 & -6250.38 & 12621.56 & 12966.80 & 0.8496 & 0.2331 \\ 
  Kullback-Leibler - 0.05 & 50.20 & -6250.88 & 12602.17 & 12889.11 & 0.8502 & 0.2371 \\ 
  Kullback-Leibler - 0.20 & 38.00 & -6262.82 & 12601.65 & 12818.86 & 0.8504 & 0.2612 \\ 
  Refined BN & 28.60 & -6286.85 & 12630.89 & 12794.37 & 0.8479 & 0.3627 \\ 
  HClust k = 2 & 16.00 & -6724.99 & 13481.97 & 13573.43 & 0.8041 & 0.2597 \\ 
   \hline
\end{tabular}
}
\caption{Mean results for \pkg{stagedtrees} algorithms over 10 replications based on the random selection of 80\% of the whole \code{selfy} dataset for the estimation of models and the remaining part for testing them. Experiments performed on a standard laptop with 
\SI{8}{\giga\byte} of RAM and an i5 \SI{3.1}{\giga\hertz} CPU.\label{tab:selfy}}
\end{table}

\begin{table}[!h]
\centering
\scalebox{0.68}{
\begin{tabular}{c@{\hspace{1cm}}|cccccccccc}
  \hline
 \multirow{2}{*}{Algorithm} & \multicolumn{10}{c}{Dataset} \\
  \cline{2-11}
  & \code{Asym} & \code{chestSim500} & \code{FallEld} & \code{monks1} & \code{PhDArticles} & \code{Pokemon} & \code{puffin} & \code{reinis} & \code{selfy} & \code{Titanic} \\ 
 \hline
\pkg{stagedtrees} BHC & 0.8490 & 0.8460 & 0.7666 & 0.9744 & 0.4164 & 0.7246 & 0.9000 & 0.8546 & 0.8491 & 0.7934 \\ 
  Bnlearn Hill-Climbing & 0.6985 & 0.6610 & 0.6942 & 0.4500 & 0.4645 & 0.7246 & 0.4385 & 0.8562 & 0.7554 & 0.6793 \\ 
  Bnlearn Tabu & 0.6985 & 0.8510 & 0.7596 & 0.4500 & 0.4754 & 0.7246 & 0.4385 & 0.8562 & 0.7804 & 0.7102 \\ 
  Logistic Model & 0.6400 & 0.8480 & 0.7667 & 0.7372 & 0.4836 & 0.7246 & 0.9385 & 0.8562 & 0.8486 & 0.7795 \\ 
  Naive Bayes Classifier & 0.6815 & 0.8480 & 0.7669 & 0.7372 & 0.4672 & 0.7246 & 0.9692 & 0.8562 & 0.8282 & 0.7752 \\ 
  Neural Network & 0.8490 & 0.8360 & 0.7668 & 1.0000 & 0.4497 & 0.7231 & 0.9538 & 0.8543 & 0.8502 & 0.7918 \\ 
  Classification Tree & 0.8490 & 0.8510 & 0.7668 & 0.7605 & 0.4530 & 0.7231 & 0.8923 & 0.8552 & 0.8516 & 0.7902 \\ 
  Random Forest & 0.8490 & 0.8490 & 0.7668 & 1.0000 & 0.4639 & 0.7231 & 0.9615 & 0.8562 & 0.8498 & 0.7925 \\ 
   \hline
\end{tabular}
}
\caption{Mean accuracies for one of the best fitting \pkg{stagedtrees} algorithm (BHC) and algorithms from the literature over 10 replications based on the random selection of 80\% of the whole dataset for model estimation and the remaining 20\% for testing. \label{table:literature}}
\end{table}

\section[A dataset analysis using stagedtrees]{A dataset analysis using stagedtrees} \label{sec:interpret}
The \code{data.frame} \code{PhDArticles} includes information regarding the number of publications of 915 PhD biochemistry students during the 1950s and 1960s 
\citep{Long1990} and it is available in the \pkg{stagedtrees} package.

The pipe operator from the \pkg{magrittr} package~\citep{magrittr} is also used. Even if it is not essential for the \pkg{stagedtrees} 
implementations and it is not one of the dependencies, the use of the pipe operator improves readability of the code and simplifies the user experience.
\begin{Schunk}
\begin{Sinput}
R> library(magrittr)
R> data("PhDArticles")
R> str(PhDArticles)
\end{Sinput}
\begin{Soutput}
'data.frame':	915 obs. of  6 variables:
 $ Articles: Factor w/ 3 levels "0","1-2",">2": 1 1 1 1 1 1 1 1 1 1 ...
 $ Gender  : Factor w/ 2 levels "male","female": 1 2 2 1 2 2 2 1 1 2 ...
 $ Kids    : Factor w/ 2 levels "yes","no": 2 2 2 1 2 1 2 1 2 2 ...
 $ Married : Factor w/ 2 levels "no","yes": 2 1 1 2 1 2 1 2 1 2 ...
 $ Mentor  : Factor w/ 3 levels "low","medium",..: 2 2 2 1 3 1 1 2 2 1 ...
 $ Prestige: Factor w/ 2 levels "low","high": 1 1 2 1 2 2 2 1 2 1 ...
\end{Soutput}
\end{Schunk}
\begin{Schunk}
\begin{Sinput}
R> bn <- bnlearn::hc(PhDArticles)
R> plot(bn)
\end{Sinput}
\end{Schunk}
\begin{Schunk}
\begin{Sinput}
R> order <- c("Gender", "Kids", "Married", "Articles")
R> bn.as.tree <- as_sevt(bn.fit(bn, data = PhDArticles), order = order)
R> plot(bn.as.tree) 
\end{Sinput}
\end{Schunk}

\begin{figure}
\begin{minipage}{0.49\textwidth}
\begin{center}
\includegraphics[scale = 1.5]{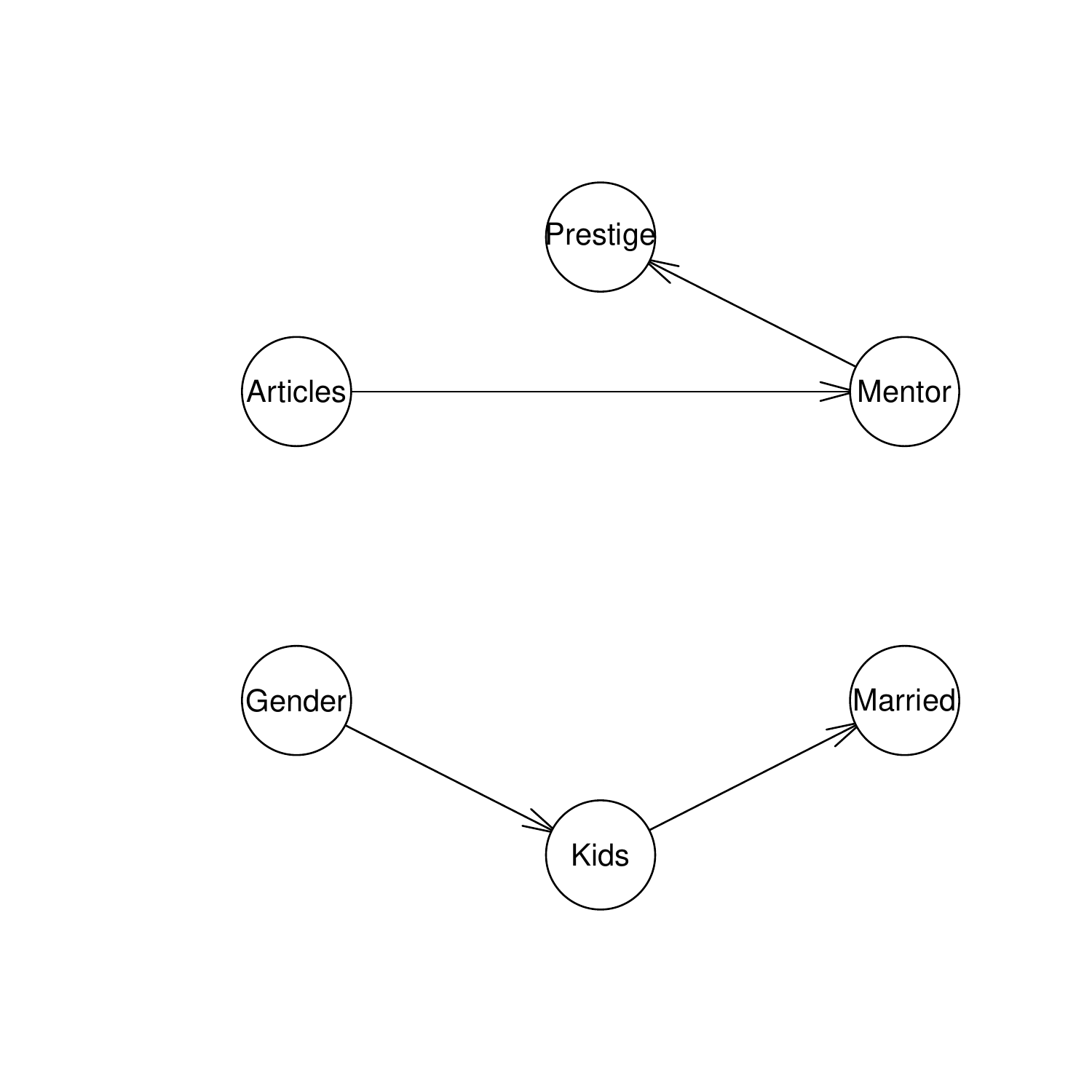} 
\end{center}
\end{minipage}
\begin{minipage}{0.49\textwidth}
\begin{center}
\includegraphics[scale = 1.5]{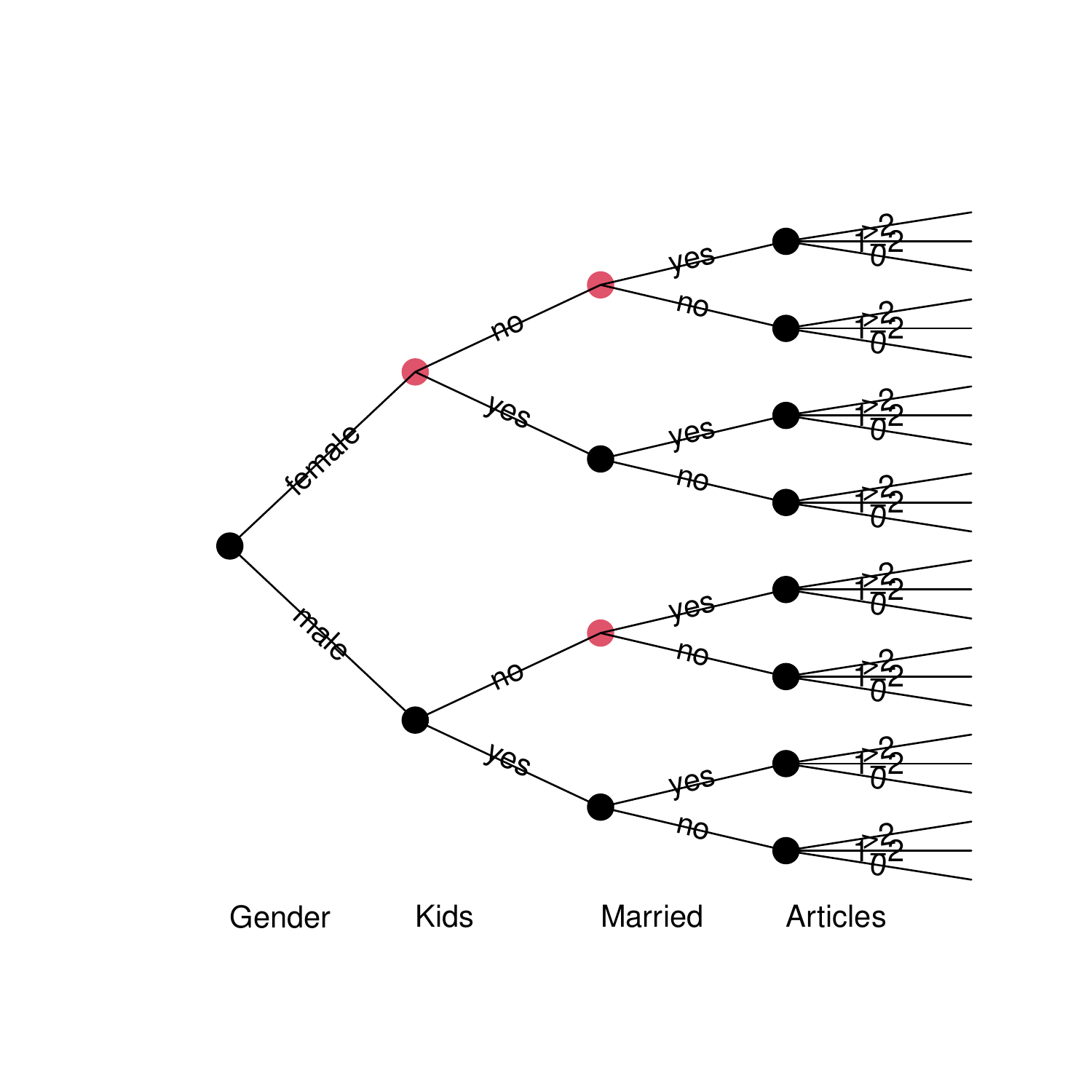} 
\end{center}
\end{minipage}
\vspace{-0.5cm}
\caption{BN model learned over  the \code{PhDArticles} dataset and 
equivalent staged tree over \code{Gender}, \code{Kids},
\code{Married} and \code{Articles}. \label{fig:bnphd}}
\end{figure}

The learned BN model in Figure \ref{fig:bnphd} left states that the
number of publications (\code{Articles}) is marginally independent 
of \code{Gender}, \code{Married} and \code{Kids} and states that 
the prestige of the University is conditionally independent of 
the number of publications of the student given the number of 
publications of the mentor.

\begin{figure}
\begin{minipage}{0.49\textwidth}
\begin{center}
\includegraphics[width=2.8in]{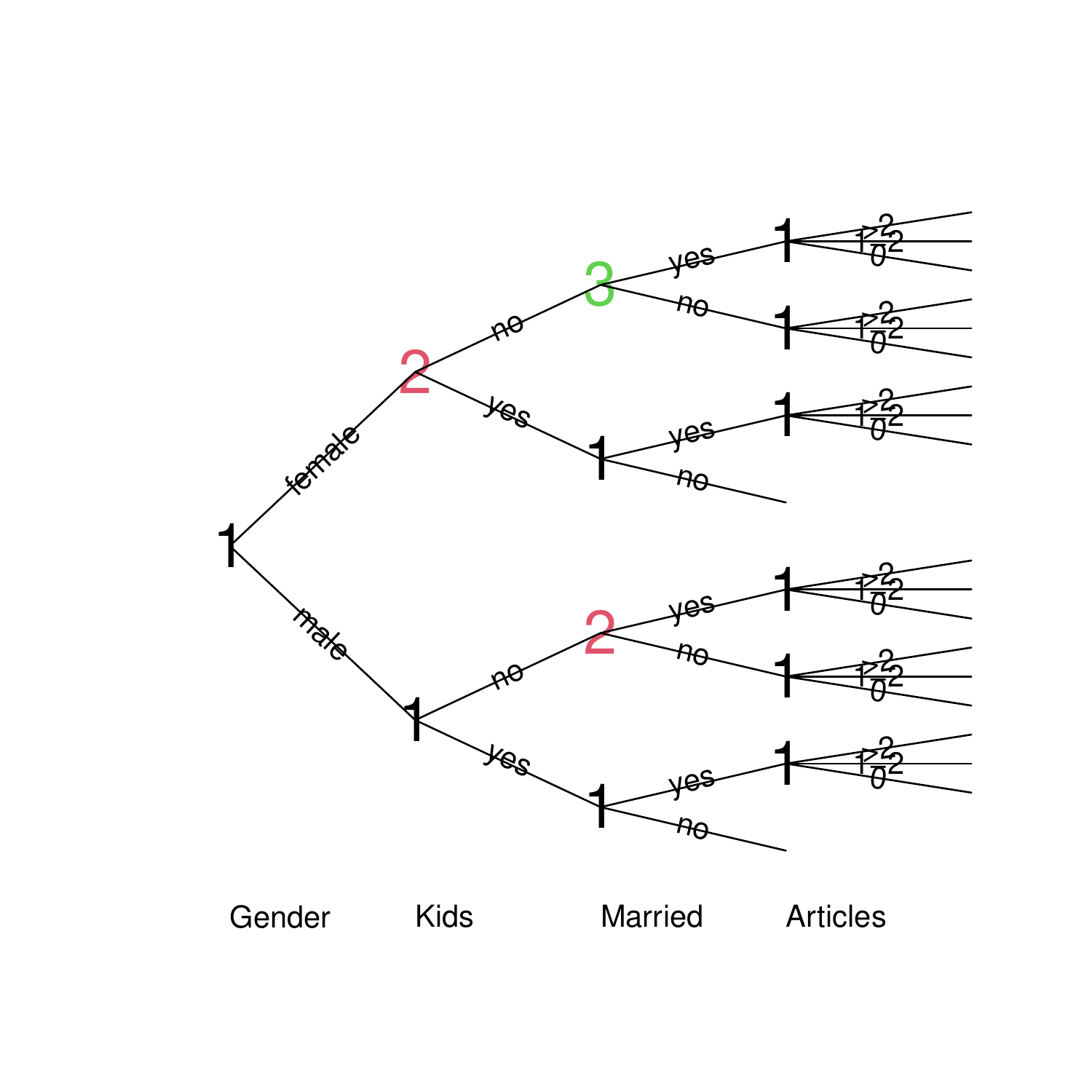} 
\end{center}
\end{minipage}
\begin{minipage}{0.49\textwidth}
\begin{center}
\includegraphics[width=2.8in]{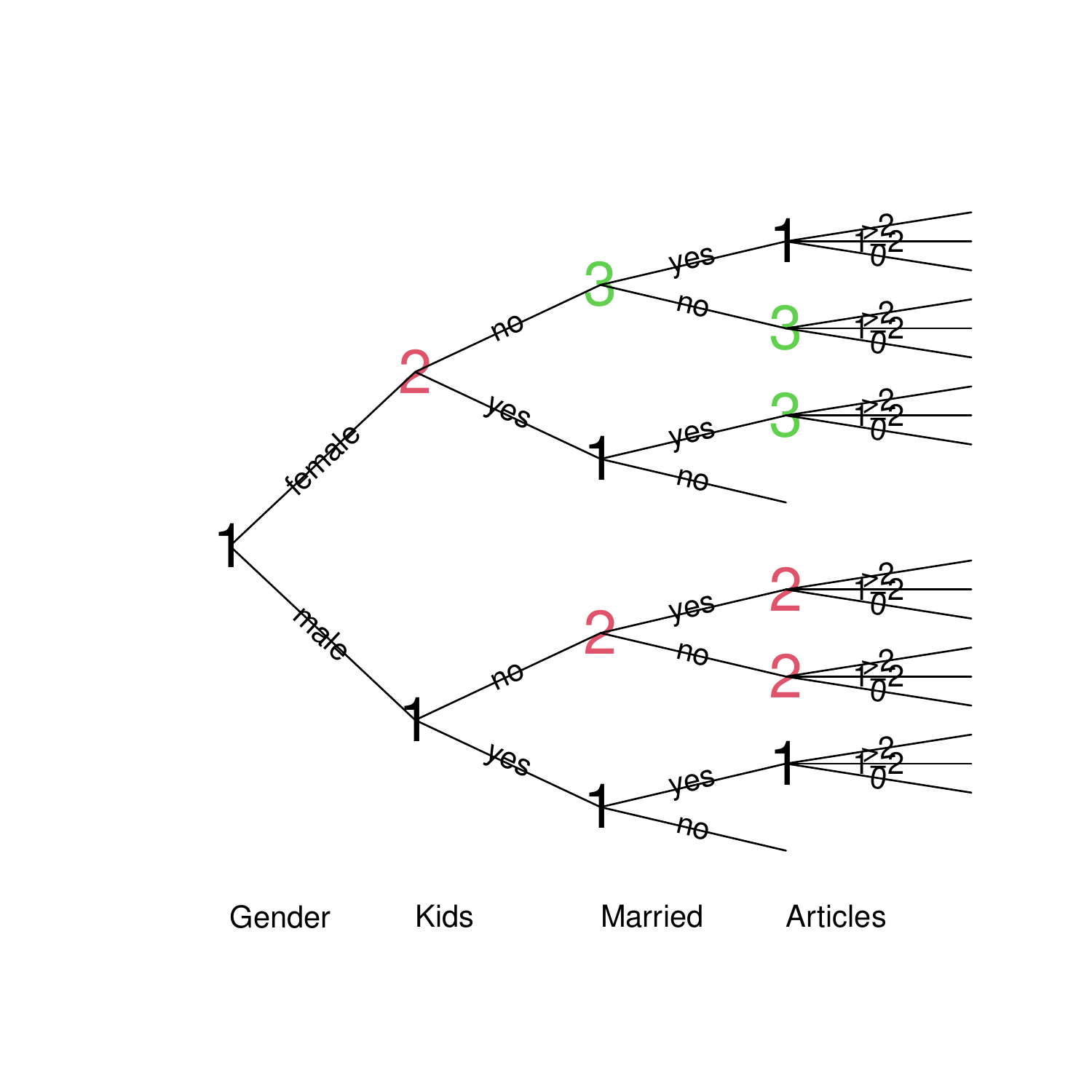} 
\end{center}
\end{minipage}
\vspace{-0.5cm}
\caption{\label{fig:phd1} Staged tree models learned over the variables \code{Gender}, \code{Kids},  \code{Married} and  \code{Articles} of \code{PhDArticles}. Left: Staged event tree \code{phd.mod1}. Right: Staged event tree \code{phd.mod2}.}
\end{figure}

The strength of the marginal independence between \code{Articles} and (\code{Gender}, \code{Kids}, \code{Married}) is investigated. 
On these four variables, staged tree models starting from the independence tree (\code{phd.mod1}) and the full tree (\code{phd.mod2}) are 
learned using the hill-climbing algorithm and are reported in Figure \ref{fig:phd1}. 
\begin{Schunk}
\begin{Sinput}
R> phd.mod1 <- PhDArticles %>% indep(order = order) %>% stages_hc
R> phd.mod2 <- PhDArticles %>% full(order = order) %>% stages_hc 
\end{Sinput}
\end{Schunk}

\begin{Schunk}
\begin{Sinput}
R> compare_stages(phd.mod1, phd.mod2, plot = TRUE, method = "stages") 
\end{Sinput}
\end{Schunk}
\begin{Schunk}
\begin{Soutput}
[1] FALSE
\end{Soutput}
\end{Schunk}

\begin{figure}
\begin{minipage}{0.49\textwidth}
\begin{center}
\includegraphics{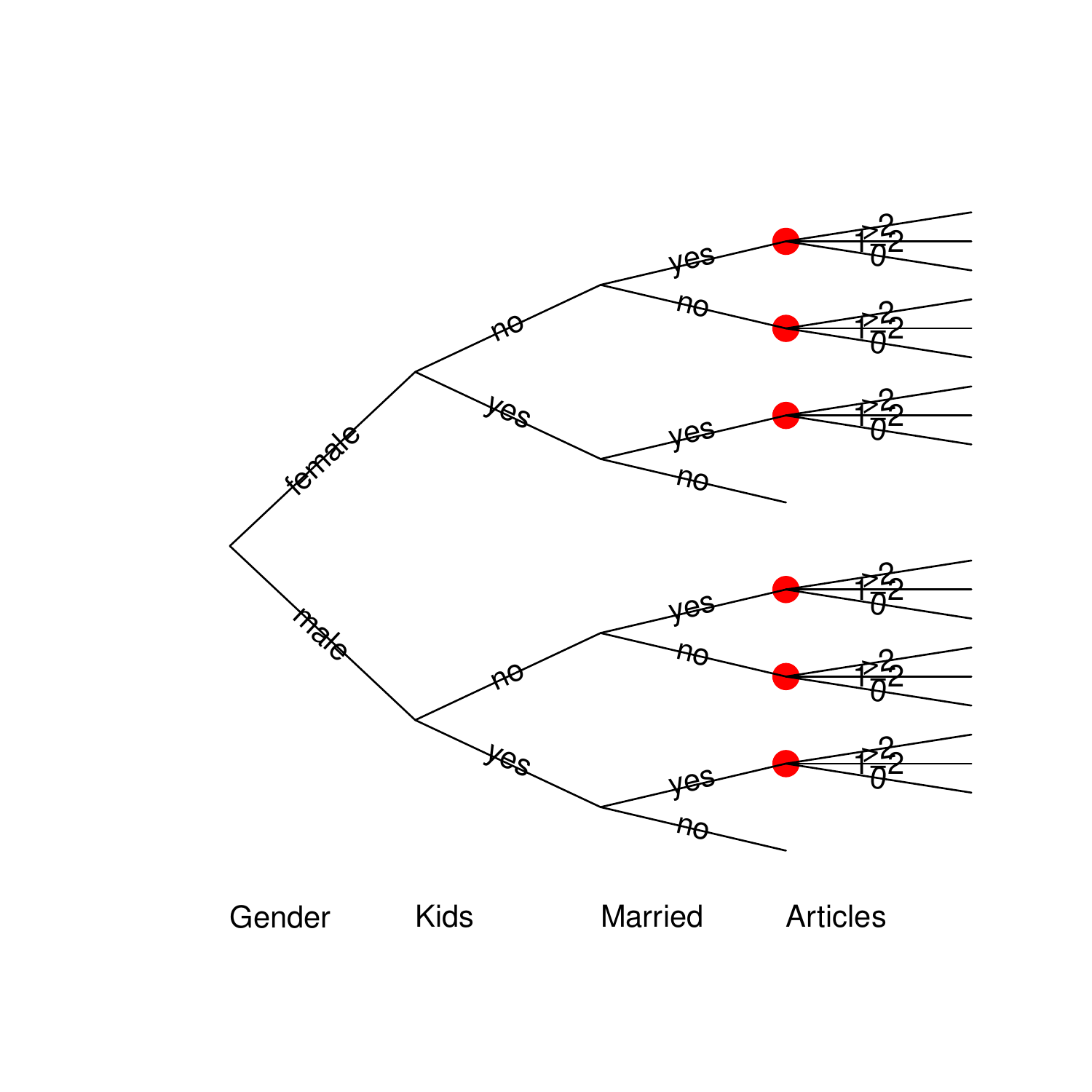} 
\end{center}
\end{minipage}
\begin{minipage}{0.49\textwidth}
\begin{center}
\includegraphics{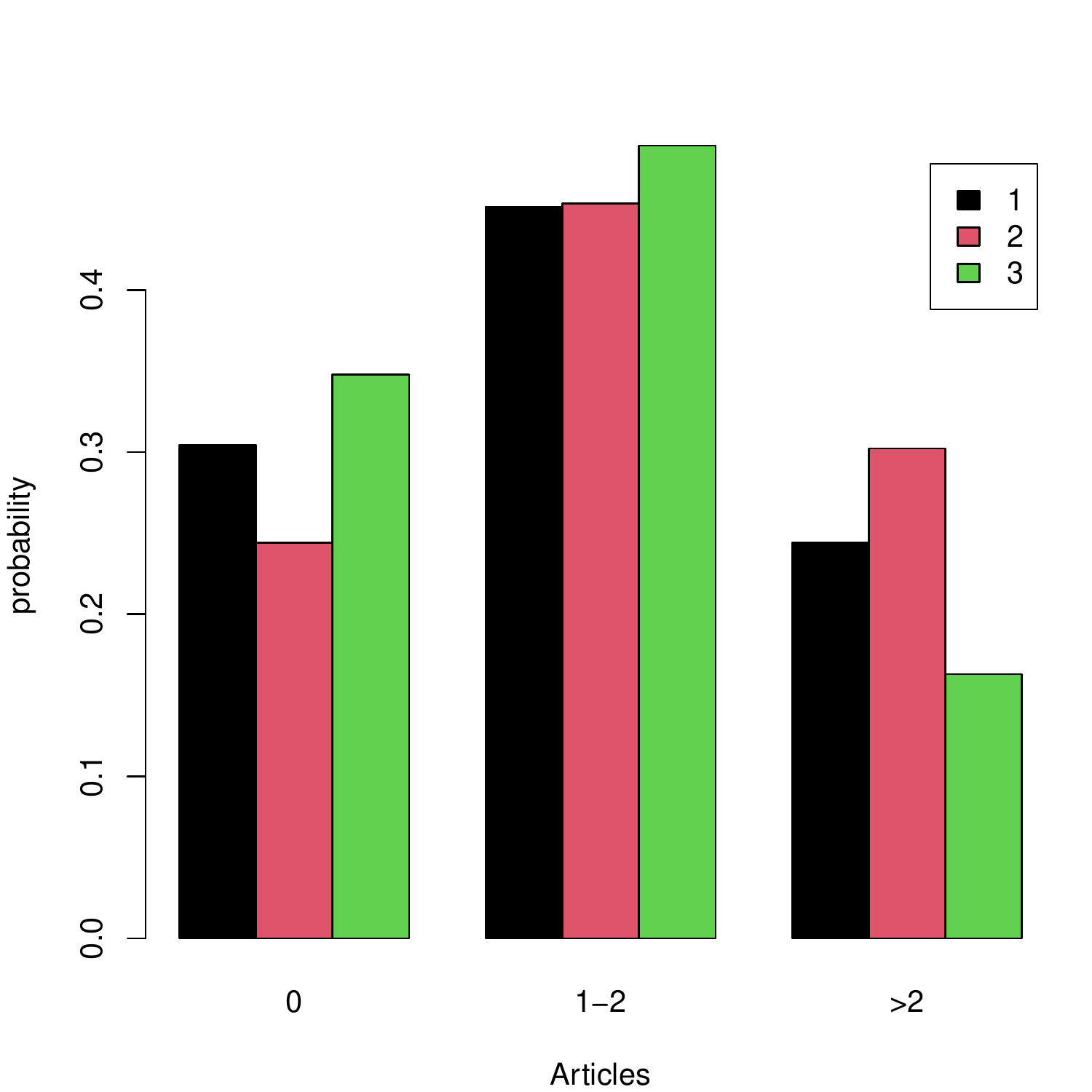} 
\end{center}
\end{minipage}
\caption{Left: Comparison between \code{phd.mod1} and \code{phd.mod2} over the variables \code{Gender}, \code{Kids},  \code{Married} and  \code{Articles} of \code{PhDArticles}.
Right: Conditional probability of \code{Articles} given \code{Gender}, \code{Kids} and \code{Married} for the stages in \code{phd.mod2}.
\label{fig:comparephd}}
\end{figure}

Investigating the estimated staging structures of the two staged trees, it is clear that for the first three variables they are exactly equal, according to the comparison depicted in Figure~\ref{fig:comparephd} left. Conversely, for the variable \code{Articles} in \code{phd.mod1} only one stage distribution is estimated and in \code{phd.mod2} three stages distributions are obtained (apart from the unobserved situations in the \code{"UNOBSERVED"} stage). 
To further explore the different conditional probabilities associated to the stages for \code{Articles} in \code{phd.mod2}, 
the \code{barplot} function can be used.
\begin{Schunk}
\begin{Sinput}
R> barplot(phd.mod2, "Articles", legend.text = TRUE, xlab = "Articles")
\end{Sinput}
\end{Schunk}
From the output in Figure~\ref{fig:comparephd} right together with the staged tree in Figure~\ref{fig:phd1} right, it can be noted that not married 
women without kids as well as married women with kids (stage $3$) have the lowest
estimated probability of a high number of articles.
The population with the highest probability of a high number of 
publications consists of men with 
no kids (stage $2$). 

A likelihood-ratio test can be carried out to 
test if the simpler \code{phd.mod1} model 
describes  the data sufficiently well compared to the more complex \code{phd.mod2}.
For complex models we can check if they are nested using the 
\code{inclusion_stages} function.
\begin{Schunk}
\begin{Sinput}
R> L1 <- logLik(phd.mod1)
R> L2 <- logLik(phd.mod2) 
R> df <- attr(L2, "df") - attr(L1, "df")
R> cat("p-value =", pchisq(2 * (L2 - L1), df = df, lower.tail = FALSE))
\end{Sinput}
\begin{Soutput}
p-value = 0.001972608
\end{Soutput}
\end{Schunk}
The small p-value obtained ($< 0.05$) confirms that the asymmetric structure
described by \code{phd.mod2} is indeed supported by the data.

\begin{figure}
\begin{minipage}{0.49\textwidth}
\begin{center}
\includegraphics{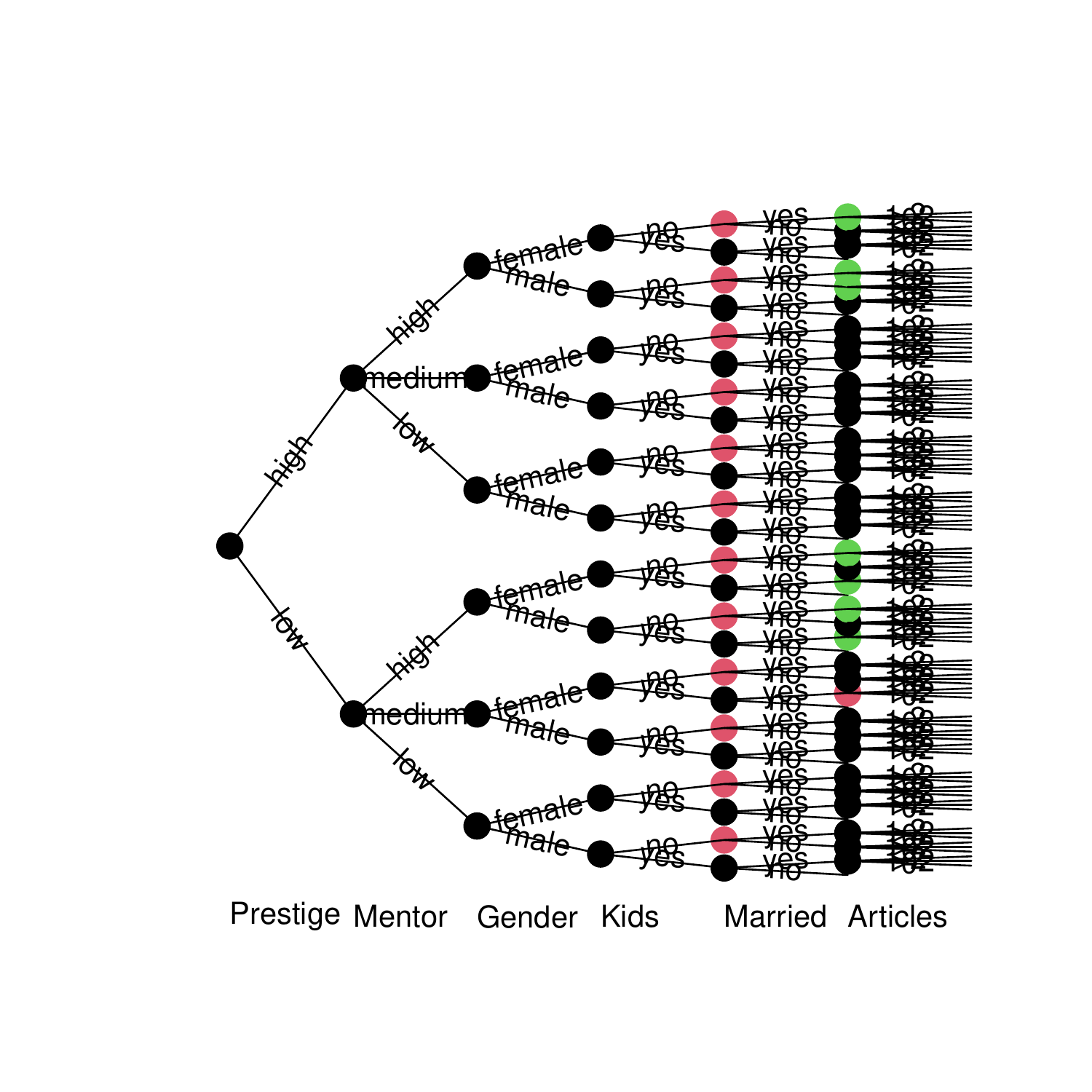} 
\end{center}
\end{minipage}
\begin{minipage}{0.49\textwidth}
\begin{center}
\includegraphics{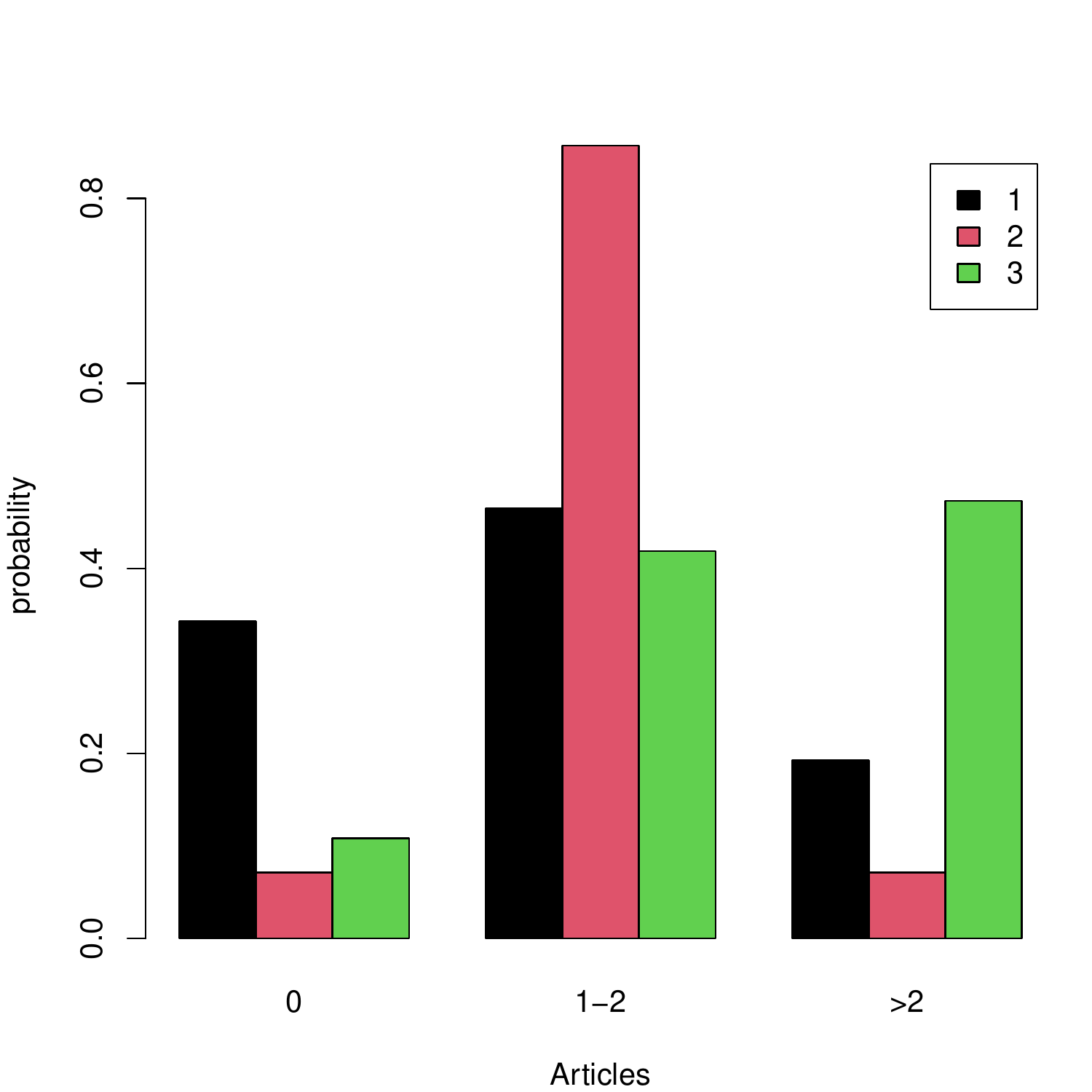} 
\end{center}
\end{minipage}
\caption{Staged tree \code{phd.all} (left) over all the variables of \code{PhDArticles}
and corresponding estimated conditional probabilities for stages related to variable \code{Articles} (right).
\label{fig:phdall}}
\end{figure}

Finally, a staged tree over all the variables in \code{PhDArticles} is
built by using the backward-joining algorithm implemented in \code{stages_bj}. 
In Figure~\ref{fig:phdall} the plot of the resulting model is displayed 
together with the barplot associated to \code{Articles} conditional 
probabilities.
\begin{Schunk}
\begin{Sinput}
R> order <- c("Prestige", "Mentor", order)
R> phd.all <- PhDArticles %>% full(order = order) %>% stages_bj(thr = 
  + 0.5) %>% stndnaming 
\end{Sinput}
\end{Schunk}

The stage with highest probability of a large number of articles (stage 3)
includes now the following paths:
\begin{Schunk}
\begin{Sinput}
R> get_path(phd.all, "Articles", "3")
\end{Sinput}
\begin{Soutput}
   Prestige Mentor Gender Kids Married
18      low   high   male  yes     yes
20      low   high   male   no     yes
22      low   high female  yes     yes
24      low   high female   no     yes
43     high   high   male   no      no
44     high   high   male   no     yes
48     high   high female   no     yes
\end{Soutput}
\end{Schunk}
So PhD students with a high number of publications all have a mentor with a high number of publications and most of them are married and with no kids.

\section[Conclusions]{Conclusions}
\pkg{stagedtrees} is an \proglang{R} package which provides a freely-available implementation of staged trees and CEGs structure. Many score functions and distances are provided for the purpose of structural learning. \pkg{stagedtrees} is designed to support users in handling categorical experimental data and analyzing the learned models to untangle complex dependence structures. It provides a set of utility functions to perform exploratory data analysis and basic inference procedures.

Only structure learning algorithms for stratified staged trees are currently implemented. The difficulty with exploring the model space of non-stratified trees lies in the exponential explosion of its size with the number of variables. Fast heuristic model search procedures are currently investigated, for instance using the maxsat approach or integer programming which have proven successful in structural learning of BNs \citep{Bartlett2017,Berg2014}.

Graphical outputs from the functions' package are produced using the \proglang{R} \pkg{graphics} package. In addition, no function is provided within the package to plot CEGs, since  no theoretical studies have been carried out yet to establish ``optimal'' representations of their underlying graph. This line of research is currently been pursued by the authors who are developing an additional \proglang{R} package to provide the user with more refined graphical functions than the ones included in \pkg{stagedtrees}.

\section*{Acknowledgments}
Motivations for the implementation of the \proglang{R} package \pkg{stagedtrees} emerged at the \textit{1st UK Workshop on Probabilistic Reasoning Using CEGs (Glasgow 2019)}. The participants are gratefully acknowledged. Professor Marco Scutari is thanked for helpful email exchanges.  The members of the SELFY project and Professor Alessandra Minello are thanked for sharing the selfy dataset.
GV was supported by a research grant (13358) from VILLUM FONDEN.







\bibliography{Bib}

\end{document}